# Deep Learning for Limit Order Books

Justin A. Sirignano[*]

May 16, 2016 [†]

**Abstract**

This paper develops a new neural network architecture for modeling spatial distributions (i.e., distributions on $\mathbb{R}^d$) which is computationally efficient and specifically designed to take advantage of the spatial structure of limit order books. The new architecture yields a low-dimensional model of price movements deep into the limit order book, allowing more effective use of information from deep in the limit order book (i.e., many levels beyond the best bid and best ask). This "spatial neural network" models the joint distribution of the state of the limit order book at a future time conditional on the current state of the limit order book. The spatial neural network outperforms status quo models such as the naive empirical model, logistic regression (with nonlinear features), and a standard neural network architecture. Both neural networks strongly outperform the logistic regression model. Due to its more effective use of information deep in the limit order book, the spatial neural network especially outperforms the standard neural network in the tail of the distribution, which is important for risk management applications. The models are trained and tested on nearly 500 U.S. stocks. Techniques from deep learning such as dropout are employed to improve performance. Due to the significant computational challenges associated with the large amount of data, models are trained with a cluster of 50 GPUs. Our data-driven approach offers new benefits for practical applications.

---

[*]jasirign@illinois.edu

[†]The author thanks the Mathematical Finance Section of the Department of Mathematics at Imperial College London for generously providing funds for computations. The author also thanks Apaar Sadhwani (Stanford University), Rob Wang (Stanford), Ilya Trubov (J.P. Morgan), Rama Cont (Imperial College London), Kay Giesecke (Stanford), Mamdouh Medhat (Cass Business School), Steven Hutt (CME), Victor DeMiguel (LBS), Xuefeng Gao (The Chinese University of Hong Kong), and David Harding (Winton Capital Management) for insightful comments.

# 1 Introduction

The "limit order book" of a stock consists of all of the active buy and sell orders at all price levels. At any point in time, it describes the known supply and demand for the stock. The prices at which the stock can be immediately bought and sold are the best ask price and best bid price of the limit order book, respectively. Modeling is challenging due to the complexity and high-dimensionality of limit order books. A limit order book has hundreds of price levels where orders may be submitted and its dynamics are nonlinear. Modeling requires the analysis of large amounts of data, which can be both statistically and computationally challenging. This paper develops a data-driven model for limit order books that addresses these challenges. We design a new deep neural network architecture for modeling spatial distributions (i.e., distributions on $\mathbb{R}^d$). This "spatial neural network" specifically takes advantage of the structure of limit order book dynamics. In our out-of-sample testing on limit order book data, the spatial neural network outperforms a standard deep neural network. Both neural networks outperform logistic regression (with nonlinear features) and the naive empirical model.

The new neural network architecture (i.e., the spatial neural network) developed in this paper has several advantages for modeling distributions on $\mathbb{R}^d$ as compared to the standard neural network architecture, including better generalization over space, lower computational expense, and the ability to take advantage of any "local spatial structure". We find statistical evidence for a particular form of local behavior in limit order books. The spatial neural network's architecture mimics this local behavior, yielding a low-dimensional model of price movements deep into the limit order book. This allows the spatial neural network to more effectively use information from deep in the limit order book (beyond the best bid and best ask prices). We use this deep spatial neural network to model the joint distribution of the best ask and best bid prices at a future time conditional on the current state of the limit order book. For the dataset considered in this paper, the spatial neural network has lower out-of-sample error, shorter training time (i.e., lower computational cost), and greater interpretability than the standard neural network.

The standard neural network can only model a distribution on a truncated region of $\mathbb{R}^d$. In contrast, the spatial neural network can model a distribution on the entire space $\mathbb{R}^d$. An important technical question that arises for the spatial neural network is whether it is a "well-posed" model for distributions on $\mathbb{R}^d$ (i.e., the distribution does not have positive mass at $\infty$). We prove that the proposed model is well-posed for several common choices of hidden units for neural networks.

We compare several approaches for modeling the joint distribution of the best ask and best bid prices at a future time conditional on the current state of the limit order book. In out-of-sample tests, neural networks strongly outperform simpler approaches such as the naive empirical model and logistic regression. The input for the logistic regression includes nonlinear features. The naive empirical model is simply the empirical distribution of the outcomes (without conditioning on the state of the limit order book). Failure to outperform the naive empirical model would imply that the state of the limit order book contains no information on future price movements. The spatial neural network outperforms the standard neural network. Both neural networks have several hidden layers and are trained using methods from deep learning such as dropout and inter-layer batch normalization. Financial firms often use relatively simple statistical models (e.g., logistic regression) in practice. The strong performance of the neural networks suggests that status quo industry risk modeling and risk management approaches can potentially benefit from adopting neural networks.

By taking advantage of the limit order book's local spatial structure, the spatial neural network can better harness information from deeper in the high-dimensional limit order book. The spatial neural network especially outperforms the standard neural network for stocks which have a stronger dependence on liquidity deeper into the limit order book (i.e., the change in the best ask/bid price has a large standard deviation). Due to its more effective use of information deep in the limit order book, the spatial neural network also strongly outperforms the standard neural network in the tail of the distribution, which is of particular interest for risk management applications.

We train and test models using limit order book data for 489 stocks (primarily from the S&P 500 and NASDAQ-100) over the time period January 1, 2014 to August 31, 2015. In total, there are roughly 50 terabytes of raw data, which is filtered to create training, validation, and test sets for the limit order book at discrete time intervals. There are substantial technical challenges to analyzing the large amounts of data and model training is computationally expensive. Distributed storage and parallel computing are used to accelerate data processing. A cluster with 50 GPUs is used to train and test the deep neural networks.

Importantly, this paper models the *joint distribution* of the best bid and best ask prices. Modeling the distribution (and not just the expected change in price) is essential for risk management applications (e.g., computing value-at-risk). Moreover, the *joint* distribution can also be very important (e.g., risk of a market-making strategy which places both bids and asks). This paper's approach could also be easily used to model the joint distribution of the best bid price, best ask price, and other limit order book features. For example, it might be of interest to model the joint distribution of the best bid price, best ask price, the best bid size, and the best ask size. Furthermore, the data-driven approach in this paper could model the distribution of the state of the *entire* limit order book at a future time conditional on the current state of the limit order book. Finally, although this paper focuses on limit order books, the spatial neural network provides benefits for any setting which requires modeling a distribution on $\mathbb{R}^d$.

In spite of the wealth of research on limit order books, there is very little literature which develops or adapts machine learning methods for modeling limit order books. Furthermore, deep learning methods have not been applied. Deep learning is arguably the best approach for data-driven modeling of the limit-order book (see Section 1.3). We review some of the current approaches to limit order book modeling in Section 1.2.

## 1.1 How does a Limit Order Book Work?

Stocks are traded via matching buy and sell orders according to an order-driven system. Orders may only be submitted at discrete price levels (determined by the "tick-size", which is $.01 in the USA). A *limit order* is a buy or sell order for a stock at a certain price. The limit order will appear in the limit order book at that price and remain there until executed or cancelled. The "limit order book" consists of all limit orders at all prices. The "bids" are the buy limit orders and the "asks" are the sell limit orders. The best ask price is the lowest sell limit order and the best bid price is the highest buy limit order. Oftentimes there can be a spread between these prices (i.e., empty price levels with no orders between the best bid and best ask prices). A *market order* is an order to immediately buy or sell the stock. A market buy order is executed at the best ask price while a market sell order is executed at the best bid price; a market order consumes some (or all) of the liquidity at the best ask or best bid price.

Figure 1 shows an example of a limit order book. The number of shares available in the limit order book to be bought/sold at $k$ discrete price levels from the best ask price is the *size at level $k$*. The bid and ask sizes measure the liquidity of the limit order book at the different price levels.

The limit order book represents the known supply and demand for the stock at different price levels.

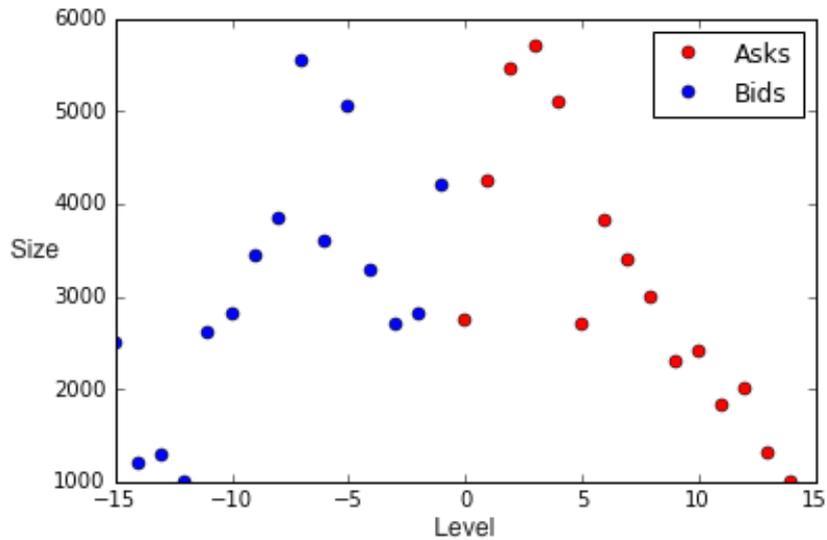

Figure 1: Bid and ask sizes for the first 15 bid and ask prices for Microsoft. Level 0 is the best ask price.

Over time, the limit order book (and with it the best ask and best bid prices) will evolve due to new limit orders, cancellations, and market orders. For practical purposes, it is of greatest interest to model the future distribution of the best ask and best bid prices given the current state of the limit order book. The best ask and best bid prices at time $t$ are the prices at which a market participant can immediately buy or sell the stock at time $t$.

The mid-price is the average of the best bid and best ask prices, and is taken as the "price" of the stock. However, it is an artificial quantity since one cannot buy or sell at the mid-price. Consequently, it is important to model both the best ask and best bid prices.

## 1.2 Related Literature

Significant research has been conducted with regards to limit order book dynamics and many theoretical models have been studied. Cont, Stoikov & Talreja (2010), Cont & Larrard (2013), Cont & Larrard (2012), Abergel & Jedidi (2013), Donier, Bonart, Mastromatteo & Bouchaud (2014), Moallemi & Saglam (2013), Maglaras, Moallemi & Zheng (2014), Carmona & Webster (2013), Carmona & Webster (2012*a*), Carmona & Webster (2012*b*), Avellaneda & Stoikov (2008), Avellaneda, Reed & Stoikov (2011), Blanchet & Chen (2013), Guo, Ruan & Zhu (2015), Gao, Dai, Dieker & Deng (2014), and others develop stochastic models for limit order book dynamics.

These theoretical models have significantly advanced our economic and mathematical understanding of the limit order book. To ensure tractability, the theoretical models impose simplifying assumptions about the dynamics of the limit order book. While such assumptions allow for the derivation and study of the mathematical and economic properties of the limit order book, they can limit the models' practical scope. Practitioners, such as financial institutions, instead often rely on statistical models (such as logistic regression) when modeling the dynamics of the limit order book.[1] Furthermore, these theoretical models do not yield tractable formulas for the joint distribution of the best ask and best bid prices at a future time horizon conditional on the current state of the limit order book; instead, the joint distribution must be estimated via simulation. Note that the joint distribution of the best ask and best bid prices is a distribution on $\mathbb{N} \times \mathbb{N} = (\ldots, -2, -1, 0, 1, 2, \ldots) \times (\ldots, -2, -1, 0, 1, 2, \ldots)$.[2] Practical applications require extremely fast predictions (milliseconds or less), making simulation undesirable.

In contrast, we develop a purely data-driven approach for modeling limit order books without any assumptions. Deep learning with neural networks is arguably the best approach to data-driven modeling of the limit order book (see Section 1.3). This paper develops a new neural network architecture which improves upon the status quo neural network architecture with lower computational cost and (in the limit order book setting) greater accuracy. The proposed neural network architecture is more interpretable than the status quo neural network architecture; this addresses a longstanding criticism of the use of neural networks for financial applications due to their lack of interpretability. Model performance is also compared against logistic regression (with nonlinear features) and the naive empirical model. By modeling the full joint distribution of the best ask and best bid prices, our model allows for the tractable calculation of a variety of quantities of interest for risk management (Value at Risk, Conditional Value at Risk, etc.). Although neural networks are costly to train, predictions can be computed quickly from the trained neural network.

There is relatively little literature on machine learning approaches to limit order books (or financial applications in general). Kercheval & Zhang (2015) use support vector machines to model whether a stock's mid-price increases, decreases, or remains the same; they do not model the actual probabilities of events, instead only predicting whether a particular event will occur. They train and test their model on only several thousand data samples. In contrast, our paper models the full joint distribution of the best bid and best ask prices, which is essential for risk management applications. We also extensively test and analyze models using many billions of data samples. Neural networks have significant advantages over support vector machines as well as other machine learning methods (see Section 1.3), with state-of-the-art results on nearly all major machine learning benchmarks. Park & Van Roy (2015) and Kearns & Nevmyvaka (2006) use reinforcement learning for optimal order execution. Reinforcement learning attempts to learn an optimal policy in an uncertain environment, while balancing the tradeoff between exploiting current knowledge and exploring new strategies.

---

[1] The theoretical models cannot be directly compared against the data-driven models in this paper due to the theoretical models' lack of tractability for computing a distribution on $\mathbb{N} \times \mathbb{N} = (\ldots, -2, -1, 0, 1, 2, \ldots) \times (\ldots, -2, -1, 0, 1, 2, \ldots)$, which is the quantity this paper is interested in. For the interested reader, we do provide a comparison in Appendix D in the much simpler case of predicting the direction of the next move of the best ask price (i.e., the probability that it moves up or down). The model from the seminal paper of Cont & Larrard (2012) is compared against logistic regression and a neural network. The data-driven models outperform the theoretical model.

[2] Modeling the future best ask and best bid prices is equivalent to modeling the number of levels by which they change.

There are also papers which apply machine learning to other areas of finance outside of limit order books. Khandani, Kim & Lo (2010), Butaru, Chen, Clark, Das & Lo (2015), Brown & Mues (2012), Baesens, Setin, Mues & Vanthienen (2003), Loterman, Brown, Martens, Mues & Baesens (2012), and Fitzpatrick & Mues (2016) use machine learning approaches for modeling consumer loan risk. Sirignano, Sadhwani & Giesecke (2016) model mortgage risk with deep neural networks. Hutchinson, Lo & Poggio (1994) use neural networks for the pricing and hedging of options. Mamaysky & Glasserman (2015) develop sentiment analysis methods for the prediction of market stress based upon news articles. Risk & Ludkovski (2015), Ludkovski (2015), and Gramacy & Ludkovski (2015) use Gaussian process regression for simulation, optimal stopping, and pricing. Dixon, Klabjan & Bang (2016) use deep neural networks for predicting financial market movements. Yang, Qiao, Beling, Scherer & Kirilenko (2015) use Gaussian process inverse reinforcement learning to identify different types of traders. Chinco, Clark-Joseph & Ye (2015) use LASSO to study cross-stock information diffusion.

## 1.3 Advantages of Neural Networks

Deep learning uses neural networks with multiple layers ("deep neural networks") in order to learn more complex nonlinear relationships from the data. Deep learning has achieved state-of-the-art results for many tasks in image classification, speech recognition, and natural language processing.[3] Neural networks are particularly well-suited for limit order books since they perform well with high-dimensional data and can capture nonlinear relationships. Neural networks also scale favorably with large amounts of data. Limit order books are high-dimensional, have nonlinear dynamics (see Section 2.1), and generate large amounts of data.

There are certainly other methods which could be applied. Decision trees, boosted trees, and random forests are also able to learn nonlinear functions. However, their disadvantage is that they divide the input space into rectangular cells while neural networks can learn arbitrary functions of the input space. The ability to learn arbitrary functions (i.e., to *generalize*) is essential for high-dimensional inputs; dividing a high-dimensional space into rectangular cells quickly suffers from the curse of dimensionality. Decision trees are also not optimized for online learning; upon the arrival of new data, typically the complete structure of the decision tree will change and it must be re-formed from scratch. In contrast, neural networks are easily trained online: the parameters can simply be updated with minibatch gradient descent. Gaussian process regression is another method which rivals neural networks in accuracy for small datasets. Unfortunately, Gaussian process regression does not scale well and becomes intractable for larger datasets. Support vector machines are shallow architectures while neural networks can have deep architectures, allowing them to learn more complex relationships; see Bengio & LeCun (2007). Support vector machines do not directly produce probabilities, instead simply providing a binary classification.[4]

## 1.4 Organization of this Paper

Section 2 describes the dataset. Section 3 presents evidence for local spatial structure in limit order books. In Section 4, different neural network architectures are analyzed for modeling spatial distributions. In particular, Section 4.3 develops the new neural network architecture for modeling spatial distributions. The deep learning approaches and the GPU computational framework used to train neural networks are explained in Section 5. Out-of-sample results for predicting the distribution of the best bid and best ask prices are reported in Section 6. Section 6 also includes analysis of the numerical results.

---

[3]For example, neural networks have achieved state-of-the-art results for the MNIST, CIFAR-10, CIFAR-100, and ImageNet datasets. Neural networks have over 99% accuracy on the MNIST dataset and 95% accuracy on the CIFAR-10 dataset. The MNIST dataset contains images of handwritten digits. The goal is to classify correctly the handwritten digits. The CIFAR-10 dataset is composed of images from ten different classes (e.g., automobiles, birds, dogs). The goal is to correctly classify an image as one of the ten classes. Mnih, Kavukcuoglu, Silver, Rusu, Veness, Bellemare, Graves, Riedmiller, Fidjeland, Ostrovski & Petersen (2015) have used deep neural networks to learn to successfully play Atari with human-level performance. Recently, Silver, Huang, Maddison, Guez, Sifre, Driessche, Schrittwieser, Antonoglou, Panneershelvam, Lanctot & Dieleman (2016) have beaten world champion players at Go using deep neural networks.

[4]There are methods to build a probability distribution on top of the support vector machine's output; a popular method is Platt scaling, which uses a logistic regression to generate a probability distribution.

## 2 The Data

We use Level III limit order book data from the NASDAQ stock exchange. For each stock, there is event-by-event data recording the current state of the limit order book. Each order submission, order cancellation, and transaction (i.e., order execution) is recorded as well as the state of the limit order book at the time of each event.[5] Both partial and full order cancellations are recorded. The times of events are reported with nanosecond decimal precision. Between events, the limit order book state does not change. The limit order book data includes the first 100 *nonzero* levels in the limit order book (50 on the ask side and 50 on the bid side). The nonzero levels are levels at which there is a nonzero bid or ask size. Thus, the dataset includes at least the first 100 levels and typically many more due to a large fraction of levels having zero sizes. At each nonzero level, the size is reported. In this paper, the "size" at a certain price refers to the number of shares available in the limit order book to be bought/sold at that price. For example, the number of shares at $k$ price levels from the best ask price is the *size at level* $k$. The sizes at ask prices are "ask sizes" and the sizes at bid prices are "bid sizes".

The data includes trading halts. During the trading halts, the limit order book is reported as unchanging. These samples are removed from the dataset for model training and testing. Trading halts can occur for various reasons, including extraordinary volatility, regulatory concerns, SEC trading suspensions, or unusual market activity indicating a technical issue or manipulation. Trading halts occur infrequently for the stocks in this paper's dataset.

The data used in this paper comes from the time period January 1, 2014 until August 31, 2015 and includes 489 stocks primarily drawn from the S&P 500 and NASDAQ-100. The large number of stocks and long time period increase the robustness of the results in this paper. Notable stocks in the dataset include Facebook, Apple, Netflix, Amazon, Amgen, Bank of America, Microsoft, Boeing, Berkshire Hathaway (Class B shares), Broadcom, and Caterpillar. A full list of stocks is provided in Appendix A. In total, the raw data is roughly 50 terabytes, which is filtered to create training, validation, and test sets for the limit order book.

---

[5]Submissions and cancellations of hidden orders are not recorded, but transactions involving hidden limit orders are recorded. Therefore, the term "limit order book" in this paper implicitly refers to the "visible limit order book".

We train and test models for two prediction cases:

1. Modeling the joint distribution of the best ask and best bid prices at time $t + \Delta t$ given the current state of the limit order book at time $t$.

2. Modeling the joint distribution of the best ask and best bid prices upon the *next price move*. The next price move is defined as the first time at which the best bid price or best ask price changes.

These two prediction cases will be referred to as Case [1] and Case [2], respectively. For Case [1], models are trained and tested specifically for $\Delta t = 1$ second, although the methodologies are of course applicable to any time horizon $\Delta t$. For Case [2], the time horizon is random. Specifically, if $\tau_1, \tau_2, \ldots$ are the times at which either the best ask price changes or the best bid price changes, we model the joint distribution of the best ask and best bid prices at time $\tau_{k+1}$ given the current state of the limit order book at time $\tau_k$. Thus, $\Delta \tau_k = \tau_{k+1} - \tau_k$ can vary widely from a fraction of a second to many seconds. Stocks which experience more frequent changes will have more data samples. For instance, AAPL has 27 million samples while FOX has 2 million samples. In Case [1], all stocks will have approximately 10 million data samples. Case [1] considers a much less volatile quantity since frequently the best bid and best ask prices do not change over a 1 second interval, while Case [2] conditions on a change occurring.[6] Case [2] is particularly interesting because the next price move is often the quantity which most directly affects the profit and loss of a strategy, position, or order execution schedule. Quantifying the magnitude of the next price move can therefore be valuable, and it has been studied in papers such as Cont & Larrard (2013), Cont & Larrard (2012), Lipton, Pesavento & Sotiropoulos (2013), and Zheng, Moulines & Abergel (2012).

Processing and storage of the raw dataset is challenging due its large size. We use distributed storage and parallel computing to store and process this data. Training complex models with many parameters (such as neural networks with multiple layers) on such a large amount of data is computationally expensive. Models are trained and tested using GPU clusters; see Section 5 for more details.

---

[6] For typical stocks, the best ask price changes only 5-15% of the time in Case [1].

## 2.1 Nonlinearity of Limit Order Books

The goal is to use neural networks to capture nonlinear relationships between the state of the limit order book and the distribution of future best bid and best ask prices. It is well-known that the limit order book has a nonlinear relationship with future price movements.[7] An example of nonlinear behavior for the stock Boeing is shown below in Figure 2; the probability that the best ask price decreases has a strong nonlinear dependence on the best ask and best bid sizes. The probability shown in the figure is the output of a neural network fitted to historical data for Boeing. The displayed relationship shows a strong dependence on supply and demand. As the best ask size increases (more selling pressure), there is an increase in the probability that the best ask price decreases. As the best bid size decreases (less buying demand), there is an increase in the probability that the best ask price decreases.

We show later that neural networks have consistently lower error than logistic regression across many different stocks. A logistic regression is the softmax of a linear function of the input while a neural network is the softmax of a nonlinear function of the input. As an input to the logistic function, we include both the original data and nonlinear features of the original data. Specifically, we include the order book imbalances at each level, which is a nonlinear function of the sizes at each level.[8] Order book imbalance (sometimes referred to as "queue imbalance") has been identified as a key driver of best bid and best ask price dynamics; see Gould & Bonart (2015), Cao, Hansch & Wang (2009), Yang & Zhu (2015), Cartea, Donnelly & Jaimungal (2015), Stoikov & Waeber (2015), and Lipton et al. (2013). The neural networks' outperformance indicates the existence of significant nonlinearity in limit order book dynamics, beyond that of the well-known order book imbalance feature.

## 2.2 Importance of modeling the *joint* distribution of best ask and best bid prices

The spread is the difference between the best bid price and best ask price. The spread is sometimes modeled as a constant. This reduces modeling the joint movements of the best bid price and best ask price to simply modeling the best ask price. Such an approach is equivalent to modeling the best bid price and best ask price as moving in lockstep: if one moves by $k$ levels, the other also moves by $k$ levels. However, statistics shown in Table 1 demonstrate that the spread frequently varies. Table 1 is based upon data from Case [1] which uses a fixed 1 second time horizon.

---

[7]For instance, see Figure 4 in Cont & Larrard (2013).

[8]Let $V_k^a$ be the size at the level ($k$ + best ask price) and let $V_k^b$ be the size at the level (best bid price $- k$). The order book imbalance at the $k$-th level is $\frac{V_k^b - V_k^a}{V_k^a + V_k^b}$. The order book imbalance ranges from $-1$ to $1$ and measures the imbalance between supply and demand for the stock.

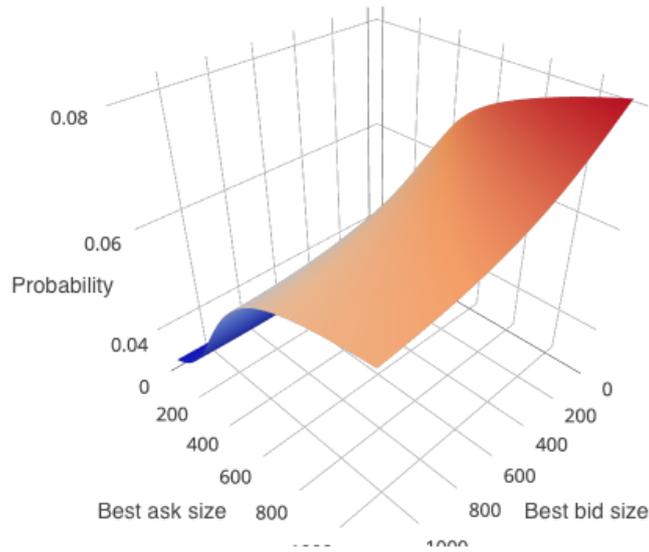

Figure 2: Probability that the best ask price for Boeing decreases for a 1-second horizon. Neural network fitted using best ask size, best bid size, and spread.

For each stock $j$ of the 489 stocks in the dataset, we empirically estimate the probability that the best bid price and best ask price change by the same amount conditional on a change in either the best bid price or best ask price occuring. That is, how frequently do the best bid and best ask prices move in lockstep? The frequency of moving in lockstep is denoted as $Z^j$ and the quantiles of $Z^1, \ldots Z^{489}$ are reported in Table 1. For the majority of stocks, the best bid price and best ask price will more than 50% of the time not move in lockstep. For half of the stocks, the best bid price and best ask price only move in lockstep 17% of the time. Table 1 also reports the (empirically estimated) probability that the best bid price changes but the best ask price does not (or vice versa) conditional on a change in either the best bid price or best ask price occurring. This frequency is denoted as $V^j$ and the quantiles of $V^1, \ldots V^{489}$ are provided in Table 1. For the vast majority of stocks, a change in one of the two will occur without a change in the other at least 25% of the time. For half of the stocks, a change in one of the two will occur without a change in the other nearly 70% of the time. In total, the first two rows of Table 1 provide strong evidence that the movements of the best ask and best bid must both be modeled. One cannot simply model the mid-price and assume that the spread is a constant number of levels. At the same time, it is also clear that the best bid and best ask prices are not independent; there is significant correlation between their movements. This highlights the importance of a model for the *joint* distribution of the best ask and best bid prices.

In addition, for each stock $j$ of the 489 stocks, we calculate the 10%, 50%, and 90% quantiles of the spread for stock $j$. We then compute the quantiles of $Q_p^1, \ldots Q_p^{489}$ where $Q_p^j$ is the $p$-percentile quantile for stock $j$. This yields insight into the sizes of the stocks' spreads in the dataset.

| Feature / Quantile (%) | 5 | 10 | 20 | 50 | 80 | 90 | 95 |
|---|---|---|---|---|---|---|---|
| $Z^j$ | .03 | .05 | .08 | .17 | .39 | .58 | .70 |
| $V^j$ | .26 | .36 | .52 | .67 | .74 | .76 | 77 |
| $Q_{10}^j$ | 1 | 1 | 1 | 1 | 2 | 3 | 6 |
| $Q_{50}^j$ | 1 | 1 | 1 | 2 | 4 | 8 | 13 |
| $Q_{90}^j$ | 1 | 1 | 2 | 5 | 10 | 18 | 28 |

Table 1: Summary statistics for the spread and the co-movement of the best bid and best ask prices.

## 3  Local spatial structure in limit order books

Limit order books exhibit some degree of local spatial structure. In Section 4.3, a new architecture for neural networks is designed which can take advantage of such local spatial structure. In this section, we provide statistical evidence for local spatial structure in limit order books.

### 3.1  Local dependence on the limit order book state

Without loss of generality, let the current best ask price at time $t$ be level $0$ and let the best ask price at time $t$ be the frame of reference for the entire limit order book. Let $Y$ be the future best ask price at time $t + \Delta t$. Conditional on $Y \geq y$ where $y > 0$, the probability that $Y > y$ strongly depends upon the ask size directly at the level $y$ of the limit order book at time $t$. The dependence on sizes at other levels is small relative to the dependence on the size at level $y$. Figure 3 demonstrates this phenomenon where the conditional movement of $Y$ depends only locally on the current limit order book state. The conditional probability that $Y > y$ given $Y \geq y$ decreases with the size at $y$.

There is some intuition regarding why the relationship in Figure 3 may hold. To reach a level $y' > y$, the sell limit orders at level $y$ must first be consumed by buy orders. The larger the ask size at level $y$, the less likely the future best ask price will reach a level $y' > y$. Since we have already conditioned on $Y \geq y$, the limit orders at levels $y' < y$ are less relevant. Similarly, the event $Y > y$ requires only that the buy orders consume the sell limit orders at $y$, so the ask sizes at levels $y' > y$ are less important.

The behavior of the best ask price as it moves upwards is analogous to a "geometric random variable" whose probability of increasing from $y$ to $y + 1$ depends upon the size at level $y$. The neural network architecture in Section 4.3 mimics this local behavior.

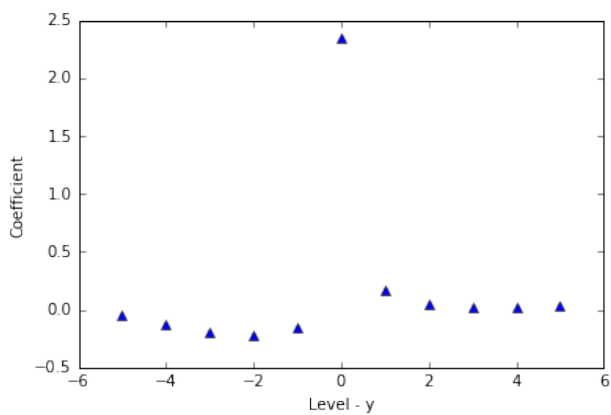

Figure 3: Coefficients from logistic regression for the probability that the future best ask price $Y$ is greater than $y$ conditional on $Y \geq y$ where $y > 0$; i.e., $\mathbb{P}[Y > y | Y \geq y] = \left(1 + \exp(b + \theta \cdot \text{factors})\right)^{-1}$. The current best ask price has been centered at 0 and the time horizon is 1 second. The plotted coefficients are coefficients for the limit order book sizes at price levels minus $y$. The conditional probability that $Y > y$ given $Y \geq y$ decreases with the ask size at $y$. The reported coefficients were fitted on the stock Amazon.

Figure (4) shows similar local structure for the stock Apple. The probability that the future best ask price $Y = y$ given that $Y \geq y > 0$ strongly depends upon the ask size at level $y$. The larger the size at level $y$, the less likely that all of the sell limit orders at that level can be consumed by the current buying demand and thus the more likely $Y$ equals $y$ (and the less likely $Y$ will move to levels greater than $y$). We also note that there is very little dependence on the best ask size in Figure (4); conditional on $Y \geq y$, the sizes at previous levels $y' < y$ in the limit order book become less influential.

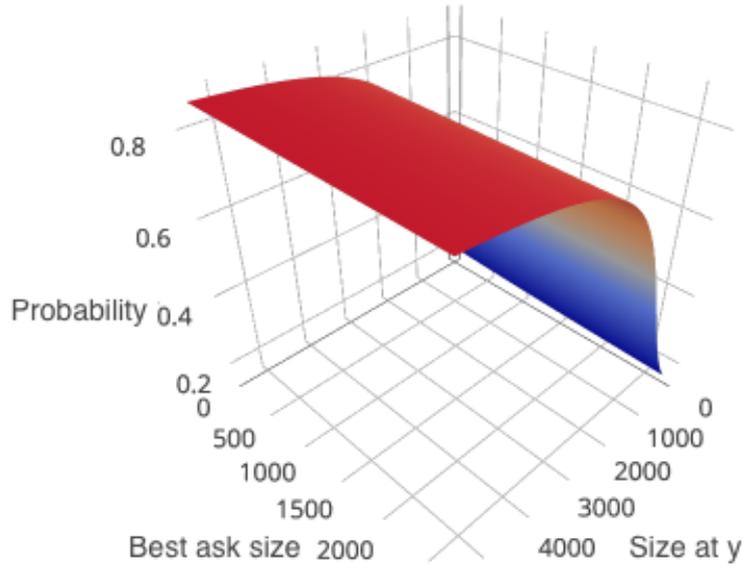

Figure 4: Probability that the future best ask price $Y = y$ given that $Y \geq y > 0$ for the stock Apple. Probability is the output of a neural network fitted to historical data for Apple and the time horizon is 1 second.

## 3.2 Statistical evidence across many stocks

Although Figure 3 is compelling, it is only one stock. A detailed analysis is now conducted across the entire dataset of 489 stocks. The results provide strong evidence for local spatial structure. For each stock, we perform a logistic regression similar to Figure 3. Specifically, let $Y$ be the best ask price at $t + \Delta t$. Without loss of generality, let 0 be the current best ask price at time $t$ and let the best ask price at time $t$ be the frame of reference for the entire limit order book. We fit a logistic regression for:

$$\mathbb{P}[Y > y | Y \geq y] = \left(1 + \exp(b + \theta \cdot (\text{size at level } y - K, \ldots, \text{size at level } y + K)\right)^{-1}, y > 0, \tag{1}$$

where $b \in \mathbb{R}$, $\theta = (\theta_{y-K}, \ldots, \theta_{y+K}) \in \mathbb{R}^{2K+1}$, and $K = 10$. The sizes are from the current limit order book state at time $t$ and are normalized. Ask sizes are given a positive sign, while bid sizes are given a negative sign.[9] Ignoring bid sizes and performing the statistical analysis solely for ask sizes yields similar results. Bid/ask sizes are from the state of the limit order book at time $t$. The time horizon $\Delta t$ is 1 second.

We fit the logistic regression (1) for each stock in the dataset, resulting in 489 different parameter fits

---
[9]The probability of an increase in the best ask price $Y$ decreases with more sell liquidity and increases with more buy liquidity, hence the opposing signs.

$\theta^1, \ldots, \theta^{489}$. Fitting is performed on the time period January 1, 2014 until May 31, 2015 (which will also be the training set used later in this paper for fitting models). For each stock $j$, the following "coefficient ratio" is calculated:

$$\text{Coefficient ratio for stock } j = \frac{\max_{y-p,\ldots,y+p} \theta_y^j}{\max_{y'=y-K,\ldots,y-p-1,y+p+1,\ldots,y+K} |\theta_{y'}^j|}. \tag{2}$$

The coefficient ratio (2) compares the local influence of levels close to $y$ versus the influence of levels farther away. $\theta_{y'}^j$ is the coefficient for the size at level $y'$ for stock $j$. The larger the magnitude of the coefficient $\theta_{y'}^j$, the greater the dependence on the size at level $y'$. If $p = 0$:

$$\text{Coefficient ratio for stock } j = \frac{\theta_y^j}{\max_{y' \neq y} |\theta_{y'}^j|}, \tag{3}$$

and the coefficient ratio measures the influence of the size locally at level $y$ versus the sizes at levels $y - K, \ldots, y - 1, y + 1, \ldots, y + K$. It also gives the direction of the dependence on the size at level $y$. If (3) is positive, then $\mathbb{P}[Y > y | Y \geq y]$ decreases as the size at level $y$ increases. Table 2 gives summary statistics for the coefficient ratio (2) across all the stocks in the dataset. There is a strong dependence on the local size at level $y$ for the majority of stocks. The sign is also positive.

| Coefficient Ratio / Quantile (%) | 5 | 10 | 20 | 50 | 80 | 90 | 95 |
|---|---|---|---|---|---|---|---|
| p = 0 | 0.84 | 1.47 | 3.38 | 6.43 | 9.89 | 13.20 | 17.69 |
| p = 1 | 1.02 | 2.31 | 5.83 | 12.83 | 19.93 | 24.71 | 30.92 |

Table 2: Summary statistics for the coefficient ratios across all stocks.

The local dependence is strongest for stocks where the change in the best ask/bid prices has a large standard deviation.[10] The standard deviation of the change in the best ask (or bid) price is a function of the price and volatility of the stock.[11] The larger the standard deviation of the change in the best ask price, the larger the dependence of the best ask price movements on liquidity deeper in the limit order book. For each stock, Figure 5 plots the coefficient ratio (3) for $p = 0$ versus the standard deviation of the change in the best ask price. Stocks where the change in the best ask price has a larger standard deviation show a stronger local dependence on the size at $y$. We observe the local dependence in the upper tail of the best ask price's distribution and the lower tail of the best bid price's distribution (i.e., when the best ask or best bid price moves *into the limit orders on their respective side of the book*).

---

[10]"The standard deviation of the change in the best ask price" is calculated by taking the standard deviation of all of the changes in the best ask price at 1 second intervals.

[11]A stock with price \$1 is much less likely to experience a move of $k$ levels than a stock with price \$500, assuming both stocks have equal volatility.

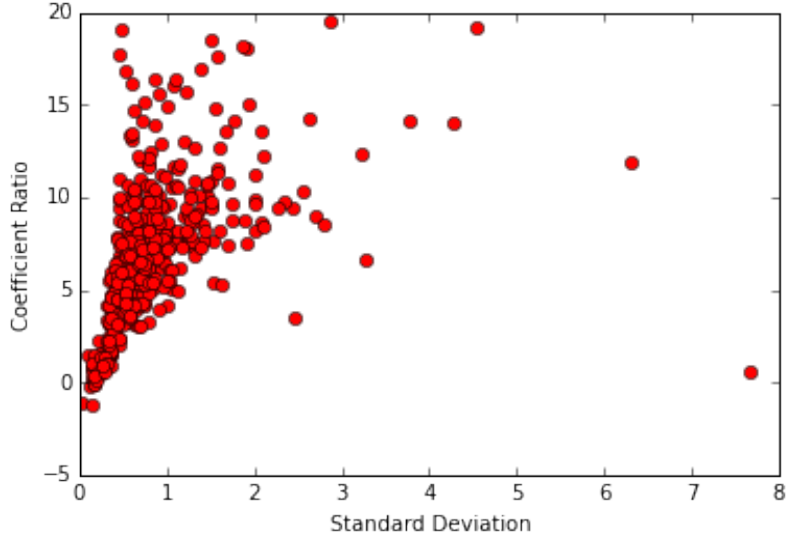

Figure 5: Plot of the coefficient ratio with $p = 0$ versus the standard deviation of changes in the best ask price for each stock. The plot has been cropped to $[0, 20] \times [0, 8]$ (some outlier data points with coefficient ratios greater than 20 are not visible in the displayed plot).

## 3.3 A Local Model for the Upper Tail of the Best Ask

The limit order book's local structure motivates a simple local model for the upper tail of the best ask price, which will later form the core of the new neural network architecture proposed in Section 4.3. Let $Y$ be the future best ask price and let $x_y$ be the size at level $y$. Then, conditional on an increase in the best ask price (i.e., $Y > 0$), we can model the magnitude of the increase as follows:

$$\mathbb{P}[Y = y | Y \geq y] = f(x_y), \quad y > 0. \tag{4}$$

(4) completely describes the distribution of the random variable $Y$ conditional on it increasing. It is analogous to modeling $Y$ as a geometric random variable, but with a non-constant probability of increasing at each step which depends locally upon the state of the limit order book. (4) mimics the local behavior described in Sections 3.1 and 3.2. Note that $\mathbb{P}[Y > y|Y \geq y]$ is simply $1 - f(x_y)$.

Although (4) has a local dependence on the state of the limit order book (only taking as an input the size at level $y$), globally the distribution of $Y$ depends upon *all of the ask sizes*. That is,

$$\mathbb{P}[Y = y|Y > 0] = f(x_y) \prod_{y'=1}^{y-1} \left(1 - f(x_{y'})\right), \quad y > 0. \tag{5}$$

An alternative to (4) would be to model $\mathbb{P}[Y = y|Y > 0]$ as a function $g(y, x_1, x_2, \ldots, x_L)$ where $L$ is the total number of levels in the limit order book. The function $g$ is far more complex than the function $f$ due to the high-dimensionality of the former's input. Furthermore, if the local behavior in (4) holds, (5) shows that $g$ will depend in a nontrivial way upon all the ask sizes as it will be the composition of many $f$ functions. Even if $f$ takes a simple form (such as a logistic regression), the global distribution (5) will be highly nonlinear, requiring $g$ to also be highly nonlinear.

The standard modeling approach would be to statistically estimate $g$. Due to the high-dimensionality of the input and the complexity of $g$, this can be challenging and may be prone to overfitting. The approach proposed in this paper is to directly estimate the local model $f$. This reduces the statistical estimation problem to estimating a low-dimensional model with a simpler functional form, potentially leading to a more accurate estimated model. Training a model to learn $f$ may also be faster than training a model to learn $g$ due to the simpler nature of $f$ in terms of both its dimension and functional form. Thus, by estimating $f$ instead of $g$, the dimension of the learning problem is significantly reduced, potentially leading to a more accurate estimated model and faster training times (i.e., lower computational cost).

The function $g$ depends nonlinearly on a high-dimensional vector, which is difficult to interpret. However, the function $f$ (and its connection to the global distribution) is straightforward to interpret. Estimating the local behavior (4) instead of the global distribution (5) therefore produces a much more interpretable model. As described in Section 3.1, there is a natural economic interpretation to the model (4); it mimics buy orders which sequentially consume the liquidities at each level. Conditional on the buy orders consuming all liquidity at levels $y' < y$, the event where it also consumes the liquidity at level $y$ depends only upon the liquidity at level $y$.

We will use an architecture similar to (4) for the neural network proposed in Section 4.3. It turns out such an architecture has additional important advantages such as computational efficiency, generalization over the output space, and the ability to model distributions on the entire positive real line.

# 4 Neural Network Architectures for Modeling Distributions on $\mathbb{R}^d$

The goal of this paper is to model a distribution on $\mathbb{R}^d$ via neural networks. Although there are many applications, we are particularly motivated by modeling the joint distribution of the best ask and best bid prices at a future time conditional on the current state of the limit order book.

In order to model a distribution on $\mathbb{R}^d$, we first discretize $\mathbb{R}^d$ into the grid $\mathcal{R}^d$ where $\mathcal{R} = \ldots, r_{-2}, r_{-1}, 0, r_1, r_2, \ldots$ and then model a distribution on the discrete space $\mathcal{R}^d$. In the context of limit order books, this discretization is exact since price levels are discrete multiples of the tick-size.

Section 4.1 reviews the standard neural network architecture for classification, which has no generalization over the space $\mathcal{R}^d$. Section 4.2 discusses a straightforward modification which allows the neural network to generalize over $\mathcal{R}^d$. However, training this neural network architecture is computationally expensive. We develop a new architecture for modeling distributions on $\mathcal{R}^d$ in Section 4.3. The proposed neural network architecture is computationally efficient, can take advantage of local spatial structure, and can be more interpretable than the aforementioned architectures. We refer to the new neural network architecture in Section 4.3 as the "spatial neural network".

## 4.1 Standard Neural Network Architecture for Classification

The basic neural network for classification is a highly nonlinear parameterized function that takes an input $x \in \mathcal{X}$ and produces a probability distribution on the finite discrete space $\mathcal{Y}$.

Given an input $x$, the output $f_{\theta,l}(x) \in \mathbb{R}^{d_l}$ of the $l$-th layer of a neural network is

$$f_{\theta,l}(x) = g^l(W_l f_{\theta,l-1}(x) + b_l), \quad l = 1, \ldots, L, \tag{6}$$

where $W_l \in \mathbb{R}^{d_l} \times \mathbb{R}^{d_{l-1}}$, $b_l \in \mathbb{R}^{d_l}$, $f_{\theta,0}(x) = x$, and $d_L = |\mathcal{Y}|$. For $l = 1, \ldots, L-1$, the nonlinear transformation $g^l(z) = \big(\sigma(z_1), \ldots, \sigma(z_{d_l})\big)$ for $z \in \mathbb{R}^{d_l}$ and $z_1, \ldots, z_{d_l} \in \mathbb{R}$. The function $\sigma$ is nonlinear; typical choices are sigmoidal functions, tanh, rectified linear units (ReLU), and clipped rectified linear units.

The function $g^L$ for the final layer $L$ is the softmax function $g$.

$$g(z) = \Big(\frac{e^{z_1}}{\sum_{i=1}^{d_L} e^{z_i}}, \ldots, \frac{e^{z_{d_L}}}{\sum_{i=1}^{d_L} e^{z_i}}\Big), \quad z \in \mathbb{R}^{d_L}. \tag{7}$$

The final output of the neural network $f_{\theta,L}(x)$ is a probability distribution on $\mathcal{Y}$ conditional on the features $x$. The parameters collectively are $\theta = (W_1, \ldots, W_L, b_1, \ldots, b_L)$, where $L$ is the number of layers in the neural network. The objective is to choose the parameters $\theta$ such that the log-likelihood $\mathcal{L}$ of the neural network's output $f_{\theta,L}$ is maximized for the data.

Let the data be $\mathcal{D} = \{(x_1, y_1), \ldots, (x_N, y_N)\}$ where $(x_n, y_n) \in \mathcal{X} \times \mathcal{Y}$. Then, the normalized log-likelihood of the data for the neural network model is

$$\mathcal{L}(\mathcal{D}) = \frac{1}{N} \sum_{n=1}^{N} \sum_{y \in \mathcal{Y}} \mathbf{1}_{y=y_n} \log f_{\theta,L}^{y}(x_n), \tag{8}$$

where $f_{\theta,L}^{y}$ is the $y$-th element of the vector $f_{\theta,L}$.

The complexity of the model is determined by both the number of layers ("depth") and the number of neurons $(d_1, \ldots, d_L)$ in each layer. Although the dividing line is somewhat arbitrary, neural networks are typically considered deep if there are three or more hidden layers ($L \geq 4$). Equation (6) describes the basic neural network achitecture, and there are several popular modifications to the architecture of the layers $1, \ldots, L-1$ (e.g., convolution neural networks). The discussion below is also applicable to these other architectures.

A potential drawback of the standard neural network for classification is that, although it generalizes over the input space $\mathcal{X}$, it does not allow for generalization over the output space $\mathcal{Y}$. As mentioned earlier, one approach to modeling a distribution on $\mathbb{R}^d$ is to discretize $\mathbb{R}^d$ into the grid $\mathcal{R}^d$ where $\mathcal{R} = \ldots, r_{-2}, r_{-1}, 0, r_1, r_2, \ldots$ and then model a distribution on the discrete space $\mathcal{R}^d$. Many problems may have some spatial structure where the event $r \in \mathcal{R}^d$ will be strongly related to the event $r' \in \mathcal{R}^d$ if $r$ and $r'$ are close in distance. A training sample at $r$ should then allow one to learn about both $r$ and $r'$. However, the standard neural network architecture for classification would regard $r$ and $r'$ as completely separate events, failing to take advantage of any available spatial structure since it has no generalization over space.

One glaring case where the standard neural network for classification fails due to a lack of generalization is when $r_k = k\Delta r$ and $\Delta r$ is small. For a dataset with $N$ samples, the fraction of grid points with at least one data sample tends to zero as $\Delta r \to 0$. Consequently, the trained neural network will predict that events at the vast majority of the grid points in $\mathcal{R}$ will occur with probability zero. A model which generalizes over the output space can avoid this pitfall. As a simple example, consider fitting a density to i.i.d. samples from a continuous random variable. In this case, there are no features (i.e., explanatory variables); formally, one can just replace the feature vector with a vector of zeros for every sample. The neural network will then trivially give the empirical measure of the observed samples, which will be zero at many grid points if $\Delta r$ is small. This is a very bad statistical estimate since the samples come from a random variable with a continuous density. Instead, one should estimate a smoothed density from the samples; the new neural network architecture developed in Section 4.3 is able to do this.

Generalizing over space is especially useful in the tails of the distribution where less data is available. The tails of the distribution, although associated with less frequent events, are important for risk analysis since they represent extreme events which can have a disproportionate impact. Generalization over space also helps to combat the curse of dimensionality. The number of grid points grows exponentially with the dimension $d$, meaning there is less data per grid point (and less data per state $y \in \mathcal{Y}$). This can be a source of overfitting.

There are other disadvantages to applying the standard neural network to modeling spatial distributions. Since $\mathcal{Y}$ is a finite discrete space, $\mathcal{R}$ must be truncated to cover only a finite region of space, which may not be desirable. Secondly, even if the bulk of the events occur in a small region of space, probabilities may still be needed at a large number of spatial points, incurring significant computational cost. For instance, even if 99% of events occur in $[0, 1]$, and the rest are uniformly spread across $[-1000, 1000]$, probabilities at all (discretized) spatial points must be calculated for each data sample during training and prediction. This incurs significant computational cost and thus slower training rates, especially in higher dimensions $d > 1$ where the number of grid points grows rapidly with $d$.

## 4.2 Straightforward modification to allow generalization

There is a straightforward modification to create a neural network which generalizes over space. This modification has been studied before; for instance, see Likas (2001). Let $f_\theta(x,y) : \mathcal{X} \times \mathcal{Y} \to \mathbb{R}$ be the unnormalized log-probability of the event $y$ conditional on the feature $x$, where $f_\theta(x,y)$ is a neural network with inputs $(x,y)$. The probability of $y$ conditional on the feature $x$ is

$$\frac{e^{f_\theta(x,y)}}{\sum_{y' \in \mathcal{Y}} e^{f_\theta(x,y')}}. \tag{9}$$

Due to the continuity of $f_\theta$, the probabilities (conditional on the feature $x$) of $y_1$ and $y_2$ will be close if the distance between $y_1$ and $y_2$ is small.

(9) can be computationally expensive. For each training sample, $f_\theta(x,y)$ and its gradient must be evaluated at every $y \in \mathcal{Y}$. If the number of training samples is $N$, this is comparable to training a standard neural network for binary classification on $N \times |\mathcal{Y}|$ training samples. For instance, if $\mathcal{Y}$ is a Cartesian grid covering $\mathbb{R}^3$ with $1,000$ grid points in each direction, $|\mathcal{Y}| = 1$ billion.

A second disadvantage is that (9) cannot model distributions on $\mathcal{R}^d$ but instead must truncate the space in order to form a finite grid.

## 4.3 A computationally efficient architecture for modeling spatial distributions

This section develops a new neural network architecture for modeling distributions on $\mathbb{R}^d$ (the "spatial neural network"). We first consider modeling a distribution on $\mathbb{R}_+ = (0, \infty)$, which is discretized into $\mathcal{R}_+ = r_1, r_2, \ldots$. Later this is extended to the more general case of $\mathcal{R}^d$. Let $f_\theta(x,y) : \mathcal{X} \times \mathbb{R} \to \mathbb{R}$ be a neural network. The distribution of a random variable $Y \in \mathcal{R}_+$ conditional on the random variable $X \in \mathcal{X}$ is completely specified by the following model:

$$\mathbb{P}[Y = y | Y \geq y, X = x] = \frac{e^{f_\theta(x,y)}}{1 + e^{f_\theta(x,y)}}. \tag{10}$$

(10) is analogous to a "geometric random variable" with a non-constant probability of increasing at each step.

The model architecture (10) is only *well-posed* if $\sum_{y \in \mathcal{R}_+} \mathbb{P}[Y = y | X = x] = 1$ for any $x$. If $\mathbb{P}[Y = $ $y\big|Y \geq y, X = x]$ decreases in $y$ without a positive lower bound, the model may not be well-posed due to probability mass escaping to $+\infty$. A trivial example is if $\mathbb{P}[Y = y|Y \geq y, X = x] = 0$ for $y \geq N_0$. Another example where $Y$ has a positive probability mass at $+\infty$ is if $\mathbb{P}[Y = y|Y \geq y, X = x] = 2^{-y-1}$ and $\mathcal{R}_+ = \mathbb{N}_+$. In both of these examples, $\sum_{y \in \mathcal{R}_+} \mathbb{P}[Y = y|X = x] < 1$.

**Theorem 4.1.** *The model architecture (10) is well-posed if the hidden units of the neural network $f_\theta(x, y)$ are bounded. The model may not be well-posed if the neural network's hidden units are rectified linear units (ReLU).*

See Appendices B and C for the proof. Theorem 4.1 covers many common choices of hidden units for neural networks, including sigmoidal, tanh, and clipped ReLU.[12] Neural networks with at least one hidden layer with bounded units (even if all the other hidden layers have unbounded units such as ReLUs) are also well-posed. If all of the hidden units in the neural network are ReLUs, the model may or may not be well-posed depending upon the form of the neural network for large $y$; see Appendix C for more details. A solution is to have at least one hidden layer of clipped ReLUs with arbitrarily large clipping values, in which case the model is well-posed.

The log-likelihood of the model (10) for a training sample $(x, y)$ is

$$\mathcal{L}(\{(x,y)\}) = \log\left(\frac{e^{f_\theta(x,y)}}{1 + e^{f_\theta(x,y)}}\right) + \sum_{y' \in \mathcal{R}_+ : y' < y} \log\left(\frac{1}{1 + e^{f_\theta(x,y')}}\right) \tag{11}$$

The architecture (10) has two advantages over (9). The first is that the neural network $f_\theta(x, y)$ and its gradient need to be evaluated at far fewer grid points. For each sample $(x, y)$, (10) only needs to be evaluated up until $y$ while (9) needs to be evaluated on the entire grid. Secondly, (10) can model the entire space $\mathcal{R}_+$; there is no need to form a truncated grid as in (9).

---

[12]A rectified linear unit is the function $\max(z, 0)$. A clipped rectified linear unit (clipped ReLU) is the function $\min(\max(z, 0), t_0)$ where $t_0 > 0$ is the clipping value.

### 4.3.1 Extension to $\mathcal{R}^d$

(10) can be extended to model distributions on $\mathcal{R}^d$. Let $Y = (Y_1, \ldots, Y_d) \in \mathcal{R}^d$ and have the conditional distribution:

$$
\begin{aligned}
\mathbb{P}[Y = (y_1, \ldots, y_d)|X = x] &= \mathbb{P}[Y_1 = y_1|X = x] \prod_{i=2}^{d} \mathbb{P}[Y_i = y_i|Y_{0:i-1} = y_{0:i-1}, X = x], \\
\mathbb{P}[Y_1 = y_1|X = x] &= g_\theta^1(x, y_1), \\
\mathbb{P}[Y_i = y_i|Y_{0:i-1} = y_{0:i-1}, X = x] &= g_\theta^i(x, y_{0:i-1}, y_i),
\end{aligned}
\tag{12}
$$

The conditional distributions $g^1, \ldots, g^d$ will be functions of neural networks, which will be specified shortly. Note that the framework (12) avoids the curse of dimensionality for large $d$ since the computational expense of the log-likelihood grows linearly with $d$:[13]

$$
\mathcal{L}(\{(x, y)\}) = \log\left(g_\theta^1(x, y_1)\right) + \sum_{i=2}^{d} \log\left(g_\theta^i(x, y_{0:i-1}, y_i)\right).
\tag{13}
$$

The conditional distribution of $Y_1$ conditional on $X$ is completely specified by:

$$
\begin{cases}
\mathbb{P}[Y_1 = y_1|Y_1 \geq y_1, X = x] = \frac{e^{f_\theta^{1,+}(x,y_1)}}{1+e^{f_\theta^{1,+}(x,y_1)}} & y_1 \geq r_1 \\
\mathbb{P}[Y_1 = z|X = x] = h_\theta^{1,z}(x) & z \in \{y_1 > 0\}, \{y_1 = 0\}, \{y_1 < 0\} \\
\mathbb{P}[Y_1 = y_1|Y_1 \leq y_1, X = x] = \frac{e^{f_\theta^{1,-}(x,y_1)}}{1+e^{f_\theta^{1,-}(x,y_1)}} & y_1 \leq r_{-1}
\end{cases}
$$

$f_\theta^{1,-} : \mathcal{X} \times \mathbb{R} \to \mathbb{R}$ and $f_\theta^{1,+} : \mathcal{X} \times \mathbb{R} \to \mathbb{R}$ are neural networks. The neural network $h_\theta^1(x)$ is a standard neural network for classification (as described in Section 4) which produces a vector of three probabilities for the events $\{y_1 > 0\}, \{y_1 = 0\}, \{y_1 < 0\}$, and $h_\theta^{1,z}(x)$ is the $z$-th vector element of $h_\theta^1(x)$. The standard neural network for classification $h_\theta^1$ is required to "stitch" together $\mathcal{R}_+$ and $\mathcal{R}_- = \ldots, r_{-2}, r_{-1}$.

Similarly, the conditional distribution of $Y_i$ conditional on $(Y_{0:i-1}, X)$ for $i \geq 2$ is completely specified

---

[13]The approach (12) can also be used with the standard neural network architecture. In order to make a fair comparison between the models, we use the approach (12) for the logistic regression, standard neural network, and spatial neural network in Section 6. Without the "dimension splitting" operation in (12), training becomes impractical for the standard neural network even in low dimensions due to the large number of grid points.

by:

$$\begin{cases} \mathbb{P}[Y_i = y_i | Y_i \geq y_i, Y_{0:i-1} = y_{0:i-1}, X = x] = \frac{e^{f_\theta^{i,+}(x,y_{0:i-1},y_i)}}{1+e^{f_\theta^{i,+}(x,y_{0:i-1},y_i)}} & y_i \geq r_1 \\ \mathbb{P}[Y_i = z | Y_{0:i-1} = y_{0:i-1}, X = x] = h_\theta^{i,z}(x, y_{0:i-1}) & z \in \{y_i > 0\}, \{y_i = 0\}, \{y_i < 0\} \\ \mathbb{P}[Y_i = y_i | Y_i \leq y_i, Y_{0:i-1} = y_{0:i-1}, X = x] = \frac{e^{f_\theta^{i,-}(x,y_{0:i-1},y_i)}}{1+e^{f_\theta^{i,-}(x,y_{0:i-1},y_i)}} & y_i \leq r_{-1} \end{cases}$$

**Example 4.2** (Limit Order Book). *Modeling the best ask and best bid prices at a future time conditional on the current state of the limit order book is equivalent to modeling the change in the best ask and best bid prices. We measure the change by the number of levels that the best ask and best bid prices move.*

$$(Y_1, Y_2) = (\text{change in best ask price}, \text{change in best bid price}) \in (\ldots, -2, -1, 0, 1, 2, \ldots)^2.$$

*The neural network $h^1$ predicts whether the best ask price will increase, decrease, or stay the same. If $h^1$ predicts that the best ask increases, $f^{1,+}$ predicts how many levels it will increase. If $h^1$ predicts that the best ask decreases, $f^{1,-}$ predicts how many levels it will decrease. $h^2$, $f^{2,+}$, and $f^{2,-}$ play similar roles for the best bid price.*

### 4.3.2 Advantages of the spatial neural network

There are several potential advantages to this proposed architecture for the spatial neural network. The model and its gradient can be evaluated at far fewer grid points in the computationally efficient architecture. Secondly, the proposed architecture can model the entire space $\mathcal{R}^d$; there is no need to form a truncated grid as in standard architectures. One disadvantage is that the architecture is composed of several neural networks instead of a single neural network. The number of neural networks grows linearly with $d$.

The proposed architecture can also take advantage of "local spatial structure", if it exists in the application setting. The spatial neural network (10) is local in nature; it models the local dynamics within a small region in space. The spatial neural network (10) can leverage a priori knowledge that conditional on $Y$ being in some region of space, the local behavior of $Y$ in that region only depends a particular subset of the values in the vector $X$. This can improve performance since it reduces the dimension of the learning problem, as described in Section 3.3. For example, if $X$ is a vector containing information at locations in $\mathbb{R}^d$ and $Y \in \mathbb{R}^d$, $Y$'s local behavior in some small region of $\mathbb{R}^d$ may only depend upon information at locations close to that small region. Such local behavior can be naturally modeled by $f_\theta(x,y)$ in (10). In the case of the limit order book, let $f_\theta(x,y) = g_\theta(m(x,y), y)$ where $g_\theta$ is a neural network and $m(x,y)$ is a map taking the vector of bid and ask sizes at all levels and outputting a smaller vector of sizes at only levels close to $y$. Although the local distribution of $Y$ conditional on $Y$ being in a particular region depends only on a subset of $X$, the global distribution of $Y$ still depends upon the entire variable $X$. In the limit order book setting where $y = (y_1, y_2) = (\text{change in best ask}, \text{change in best bid})$, the map $m(x,y)$ would output a vector of bid and ask sizes at levels close to $(y_1 + \text{current best ask}, y_2 + \text{current best bid})$.

By significantly reducing the dimension of the input space, the spatial neural network becomes much more interpretable than the standard neural network. In the case of limit order books, there is also a natural economic interpretation of the local behavior that the spatial neural network models; see Section 3.

The spatial neural network's compoutational cost only grows linearly with the dimension $d$ due to the dimension splitting in (12). However, the trick (12) is not unique to the spatial neural network and can also be used with the standard neural network. Since the computational cost only grows linearly with $d$, it would be feasible to model the distribution of the state of the *entire* limit order book using neural networks.

## 4.4 Other approaches to modeling spatial distributions

Another approach to modeling spatial distributions is to use Gaussian mixtures and model the parameters (means, covariances, and mixture weights) with neural networks. This produces a continuous distribution on $\mathbb{R}^d$. Various frameworks combining Gaussian mixtures with neural networks have been proposed by Variani, McDermott & Helgold (2015), Demuynck & Triefenbach (2013), Paulik (2013), van den Oord & Schrauwen (2014), Yu & Seltzer (2011), Sainath, Kingsbury & Ramabhadran (2012), Deng & Chen (2014), and others.

Gaussian mixture models are not suitable for the limiting order book setting. The distribution of the best ask and best bid prices does not have a density since the best ask and best bid prices take values at discrete levels. A Gaussian mixture model would converge during training to a mixture of Gaussians with zero variances, meaning there's no advantage over the standard neural network architecture which models distributions on a discrete space. Numerical difficulties may also emerge as the variances become small. In other applications outside of limit order books, Gaussian mixture models may have an advantage over the "spatial neural network" developed in this paper when the disribution is smooth and its tails are close to Gaussian. It should also be emphasized that Gaussian mixture models produce an actual density while the architectures in this paper require first discretizing space.

The spatial neural network proposed in this paper has some other advantages over Gaussian mixture models. Gaussian mixture models may require a large number of Gaussians to model sharp (or discontinuous) changes. A large number of Gaussians may also be needed if the tail is not Gaussian. Gaussian mixtures cannot model local spatial structure. Finally, as mentioned above, Gaussian mixtures are not suitable for distributions with delta functions.

## 5 Model Training

The neural networks are trained using approaches from deep learning, which we describe in Section 5.1. The computational implementation using GPU clusters is outlined in Section 5.2. The division of the dataset into training, validation, and test sets is specified in Section 5.3. Model hyperparameters are provided in Section 5.4.

### 5.1 Deep Learning

We use $4$ layers for the neural networks. Neural networks with 3 or more hidden layers are referred to as "deep neural networks". Deep neural networks are able to extract richer and more complex nonlinear relationships than "shallow" neural networks. Each additional layer extracts increasingly nonlinear features from the data. Early layers pick up simpler features while later layers will build upon these simple features to produce more complex features. Recent research has developed many new methods for training deep neural networks, and we employ several of these techniques. We use dropout to prevent overfitting (Srivastava, Hinton, Krizhevsky, Sutskever & Salakhutdinov 2014). Batch normalization is used between each hidden layer to prevent internal covariate shift (Ioffe & Szegedy 2015). The RMSProp algorithm is used for training (Graves 2013). RMSProp is similar to stochastic gradient descent with momentum but it normalizes the gradient by a running average of the past gradients. We use an adaptive learning rate where the learning rate is decreased by a constant factor whenever the training error increases over a training epoch. Early stopping via a validation set is imposed to reduce overfitting (Bengio 2012). We also include an $\ell^2$ penalty when training in order to reduce overfitting. Although ReLU units have often produced the best performance for deep neural networks, it may be preferable in the limit order book setting to use hidden units which are bounded (e.g., clipped ReLU, sigmoidal, or tanh). The bid and ask sizes are unbounded, and a small fraction have very large values. These outlier values can cause undesirably large gradient steps.

In order to make the comparison between the standard neural network architecture and the spatial neural network as fair as possible, we apply the methods above in exactly the same manner when training both of the neural networks. More discussion is provided in Section 6.

## 5.2 Computational Approach

Due to the size of the dataset and the large number of parameters in the neural networks, model training is computationally expensive. To address this, we use a cluster with 50 GPUs to train the models. Accessing and processing data is accelerated via distributed storage. Pre-processing of data is performed using parallelization over 150 vCPUs on multiple high-performance multi-core processors.

Model training is parallelized across 50 GPUs. Each GPU itself has 1,500 CUDA cores. GPUs allow massive parallelization via the large number of cores and have become the preferred approach for neural network training. We also use NVIDIA's cuDNN library, which is a highly optimized library of primitives for training deep neural networks on GPUs. In total, model training takes over 3,000 "GPU node hours"; i.e., it would take a single GPU node 3,000 hours to train all the models. Training models on the GPU is 10 times faster than training with a CPU, meaning that training on a single (non-GPU) node would take *years* to train all the models.

Filtering the original raw data to create datasets for model training is itself very challenging. The original dataset contains roughly 50 terabytes of raw data. Data is distributed across multiple storage devices. Data processing is parallelized across 5 compute-optimized Intel Xeon E5-2666 v3 Haswell processors. Each processor has 36 vCPUs, for a total of 180 vCPUs.

In Case [1] (fixed time horizon of 1 s), each stock has roughly 10 million samples. Over the entire dataset of 489 stocks, this makes for 5 billion data samples in total. In Case [2] (random time horizon at which the next change in the best ask or best bid prices occurs), each stock has on average 5 million samples.

Over the entire dataset, this amounts to 2.5 billion data samples. Each sample contains a vector of length 200, recording the state of the limit order book across the first 50 bid and ask *nonzero* levels.

## 5.3 Division of Data into Training, Validation, and Test Sets

The data is divided into three sets. The test set is all data from June 1, 2015 to August 31, 2015. The training data is composed of 95% of the data from January 1, 2014 to May 31, 2015 (drawn at random). The validation set is the remaining 5% of the data from January 1, 2014 to May 31, 2015. Models are trained and tested separately on each stock; i.e., a new randomly initialized model is trained for each stock.

## 5.4 Hyperparameters

Out-of-sample results are reported in Section 6. Both the standard neural network and spatial neural network have 4 layers. The deep learning techniques discussed in Section 5.1 are applied in the same manner to the standard neural network and spatial neural network. In addition, all models use the same batchsize, initial learning rate, momentum, $\ell^2$ penalty, and dropout rate. All models are trained using the RMSProp algorithm. The initial parameters are randomly initialized for all models. Although the results reported in Section 6 are for a single set of hyperparameter choices for the batchsize, initial learning rate, momentum, and dropout rate, we did test a wide range of values for these hyperparameters on a small subset of stocks and found that results are robust to the choice of these hyperparameters. Neurons in the hidden layers are the tanh function. We use 250 neurons per hidden layer for the standard neural network. The standard neural network needs a large number of neurons for peak performance, with smaller networks (e.g., 50 neurons per hidden layer) not performing as well. The spatial neural network can perform well with relatively few neurons, and we use 50 neurons per hidden layer to save computational time.[14] All models are trained for 75 epochs.[15] After each epoch, the training data is randomly rescrambled for all models.[16] A new random initialization of the model parameters is used at the start of training for each stock. A form of early stopping is applied to the training of all models: the validation error and fitted model are recorded after each epoch and the model fit with the lowest validation error is selected from the sequence of model fits. Inputs to the logistic regression, standard neural network, and spatial neural network only include the sizes (and, in the case of logistic regression, order book imbalances) at each level of the limit order book and not the actual prices associated with those levels (e.g., the best ask price, best bid price, and mid-price are not included). The input to the spatial neural network includes both sizes for levels close to the origin as well as sizes for "local" levels near to $y$.

---

[14]The standard neural network has over 170,000 parameters while the spatial neural network has 20,000 parameters. Despite the standard neural network being a much more complex model than the spatial neural network, the spatial neural network outperforms the standard neural network. Note that this is not due to the standard neural network overfitting; as mentioned above, we found that smaller sizes (e.g., 50 neurons per hidden layer) of the standard neural network performed worse. The spatial neural network can be a much less complex model since the dimension of the learning problem has been considerably reduced and the limit order book's nonlinear behavior in the tails (i.e., the local spatial structure) has already been embedded into the model. Even smaller sizes (e.g., 10 or 25 neurons per hidden layer) for the spatial neural network can still perform well, with only a moderate decrease in performance. The smaller size of the spatial neural network means that it can be trained more quickly than the standard neural network.

[15]An epoch is a complete pass through the entire training set.

# 6 Out-of-sample Results

Out-of-sample results for Case [1] and Case [2] are reported in Sections 6.1 and 6.2, respectively. Case [1] is a fixed time horizon of 1 second and Case [2] is a random time horizon at the next price move. (See Section 2 for a detailed description.) For both Cases [1] and [2], out-of-sample performance is reported for the marginal distribution of the best ask price as well as the joint distribution of the best ask and best bid prices. The spatial neural network outperforms the standard neural network with lower error and higher accuracy. The "error" reported is the cross-entropy error, which is equivalent to the negative log-likelihood. In Section 6.3, we compare model performance in the tail of the distribution. The spatial neural network significantly outperforms the standard neural network in the tail of the distribution, which is important for risk management applications. Section 6.4 discusses the error versus computational cost for the neural networks.

Neural network results are also compared against baseline models. The first baseline model is the naive empirical model, which is simply the naive empirical distribution from the training set. If models do not have lower errors than the naive empirical model, then the limit order book contains no information on future movements of the best ask and best bid prices. The second baseline model is a logistic regression whose input includes nonlinear features (the order book imbalances). The logistic regression's input also includes the sizes at the different levels and the spread. If the neural networks have lower errors than the logistic regression, this indicates that limit order book dynamics have significant nonlinearity beyond the nonlinearity of the order book imbalance feature. Both of the neural networks strongly outperform these baseline models. The baseline models also perform poorly for modeling the tail of the distribution as compared to the neural networks (see Section 6.3).

---

[16]Note that we have randomly scrambled data over all time periods. This is important in order not to bias training towards any particular time period.

Modeling the best ask and best bid prices at a future time is equivalent to modeling the change in the best ask and best bid prices; see Example 4.2. We measure the change by the number of levels that the best ask and best bid prices move. Since the standard neural network, logistic regression, and naive model cannot model the entire real line, $\mathcal{R}$ is truncated to $-50, -49, \ldots, 49, 50$ for the purpose of model comparison.

The dimension splitting trick (12) is applied to the standard neural network and logistic regression when modeling the joint distribution of the best ask and best bid prices. This is done for two reasons. First, this makes the other models consistent with the spatial neural network and allows for a fair comparison of performances. Secondly, without using (12), the number of output states becomes $|\mathcal{R}|^2 = 10,201$ and the standard neural network's convergence during training is so slow that this approach becomes impractical even with the large amount of computational resources. Similarly, the logistic regression training also becomes very slow.

All models are trained and tested separately on each stock; i.e., models $\mathcal{M}_i^1, \mathcal{M}_i^2, \ldots, \mathcal{M}_i^{489}$ are trained where $i \in \{$naive empirical model, logistic regression, standard neural network, spatial neural network$\}$. The model $\mathcal{M}_i^j$ is trained only on the training set for stock $j$ and $\mathcal{E}_i^j$ is the out-of-sample error of the model $\mathcal{M}_i^j$ on the test set for stock $j$. A new random initialization of the model parameters is used for each stock at the beginning of training.

The spatial neural network outperforms the standard neural network on 94% of stocks in Case [1] and 97% of stocks in Case [2]. The average decrease in error for modeling the joint distribution is $0.6\%$ and $3.5\%$ in Case [1] and Case [2], respectively. The spatial neural network outperforms the logistic regression and naive empirical model on 100% of the stocks. The standard neural network outperforms logistic regression and the naive empirical model on nearly 100% of the stocks. In Case [1], the neural networks have on average 10% lower error for modeling the joint distribution as compared to the logistic regression. In Case [2], the neural networks have on average 20% lower error for modeling the joint distribution as compared to the logistic regression. A detailed report of model performances is provided in Sections 6.1 and 6.2. Section 6.1 compares model performances for Case [1]. Section 6.2 compares model performances for Case [2].

The logistic regression fairs very poorly when used to model the joint distribution in Case [1]. This is because the conditional distribution of the best bid price given the change in the best ask price is a nonlinear function. Note that the logistic regression does not have this disadvantage when modeling the joint distribution in Case [2] since either the best ask price or the best bid price changes (but not both at once). We have embedded this a priori knowledge into the models for Case [2].[17]

The spatial neural network more strongly outperforms the standard neural network in Case [2]. Case [2] is particularly interesting because the next price move is often the quantity which most directly affects the profit and loss of a strategy, position, or order execution schedule. Quantifying the magnitude of the next price move can therefore be valuable. The spatial neural network performs better relative to the standard neural network in Case [2] due to Case [2] conditioning on a change in the best ask price or best bid price. As described in Section 3, the spatial neural network takes advantage of local spatial structure in the tails of the distribution (in particular when the bet ask price increases or the best bid price decreases). For typical stocks, the best ask price changes only 5-15% of the time in Case [1]. Consequently, the advantage of the spatial neural network is not applicable for the majority of samples. Thus, even though the spatial neural network strongly outperforms conditional on a movement, the overall error in Case [1] only decreases modestly since a price move occurs only a small percentage of the time. Case [2] conditions on such a movement occuring, so the outperformance is larger. Section 6.3 examines model performance in the tail of the distribution for Case [1] and finds that the spatial neural network strongly outperforms the standard neural network in the tail, which matches both the statistical evidence in Section 3 and the performance results for Case [2].

The advantages of the spatial neural network in the tail of the distribution can be useful for risk management purposes. Risk analysis is primarily concerned with rare events. For instance, a risk manager might ask conditional on a price change at a certain time horizon, how large might the price move be? Or a risk manager might ask how large will the next price move be? Sections 6.2 and 6.3 show that the spatial neural network provides a large improvement in performance in the tail of the distribution. Besides the cross-entropy error, Section 6.3 also reports the accuracy of the different models in the tail of the distribution. Accuracy is another metric which can be used to evaluate model performance. Model accuracy is the percentage of the time where the model correctly predicts the outcome.

---

[17]In Case [2], given the best ask price changes, the conditional log-likelihood of the best bid price not changing is 0.

Figure 6 compares the out-of-sample performances of the spatial neural network and the standard neural network in the upper tail of the distribution for Case [1]. Here, the "upper tail of the distribution" is the distribution conditional on the best ask price increasing. The outperformance of the spatial neural network relative to the standard neural network increases with the standard deviation of the change in the best ask price. This matches the statistical behavior found in Section 3 (in particular, see Figure 5) where the local dependence was stronger for stocks with larger standard deviations. Stocks with larger standard deviations have a stronger dependence on liquidity deeper in the limit order book. Figure 7 compares the out-of-sample accuracies of the spatial neural network and the standard neural network in the upper tail of the distribution for Case [1]. Again, the outperformance increases with the standard deviation of the change in the best ask price.

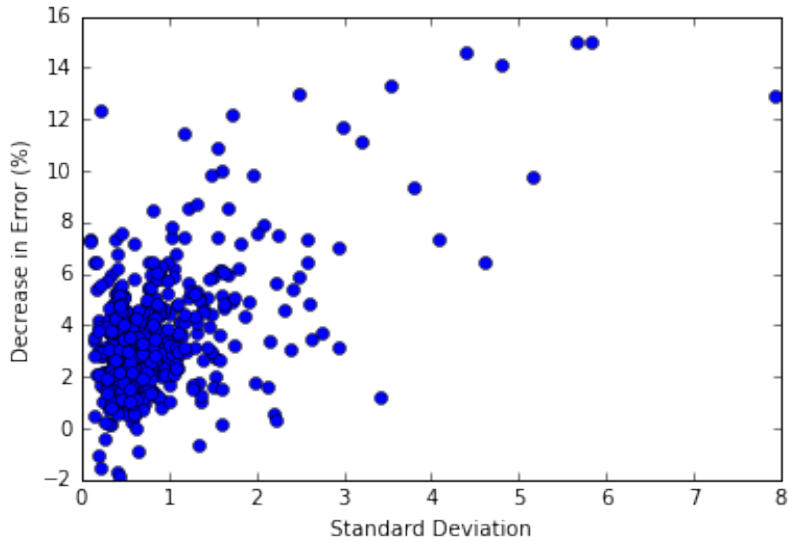

Figure 6: Decrease in out-of-sample error of the spatial neural network versus the standard neural network plotted against the standard deviation of the change in the best ask price. Decrease in error for stock $j$ is $\frac{\varepsilon^j_{\text{Standard Neural Network}} - \varepsilon^j_{\text{Spatial Neural Network}}}{\varepsilon^j_{\text{Standard Neural Network}}} \times 100\%$. Results are for the marginal distribution of the best ask price at a 1 second time horizon *conditional on the best ask price increasing*.

Finally, we find that the spatial neural network's outperformance can largely be attributed to taking advantage of the local spatial structure described in Section 3. To take advantage of the local spatial structure, the spatial neural network requires as an input the "local" state of the limit order book (i.e., the sizes for level $y$ and nearby levels). We tested the spatial neural network without these local inputs and performance decreased significantly.

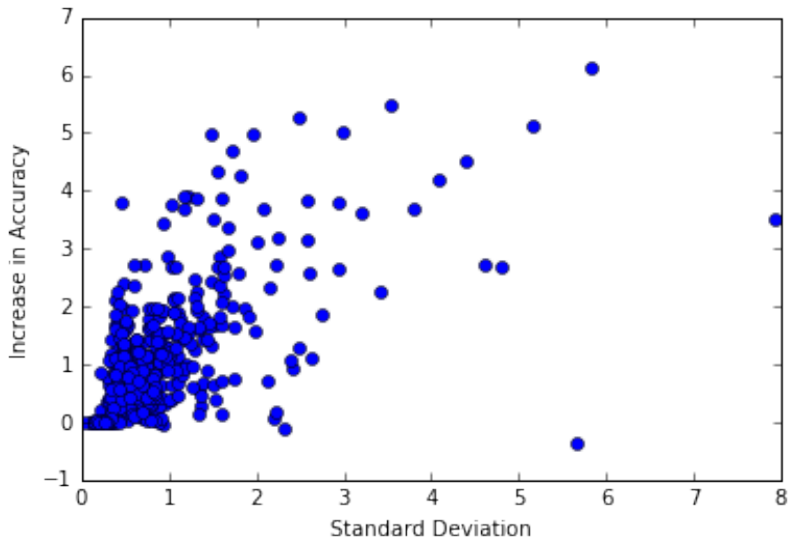

Figure 7: Out-of-sample accuracy of the spatial neural network minus the out-of-sample accuracy of the standard neural network plotted against the standard deviation of the change in the best ask price. Results are for the marginal distribution of the best ask price at a 1 second time horizon *conditional on the best ask price increasing*.

## 6.1   Case [1]: Fixed Time Horizon of 1 second

Out-of-sample performance of the models is reported for the prediction case of a fixed time horizon of 1 second. The models seek to predict the joint distribution of the best ask and best bid prices at time $t + \Delta t$ given the current state of the limit order book at time $t$.

Tables 3 and 4 compare the out-of-sample performance of the different models for the marginal distribution of the best ask price for Case [1]. Tables 5 and 6 compare the out-of-sample performance of the different models for the *joint distribution* of the best ask and best bid prices for Case [1]. The neural networks consistently have lower error than the naive empirical model and the logistic regression. The spatial neural network consistently has lower error than the standard neural network.

| Model 1/Model 2 | Naive empirical model | Logistic Reg. | Neural Net. | Spatial Neural Net. |
|---|---|---|---|---|
| Naive empirical model | NA | 4/489 | 1/489 | 0/489 |
| Logistic Reg. | 485/489 | NA | 2/489 | 0/489 |
| Neural Net. | 488/489 | 487/489 | NA | 31/489 |
| Spatial Neural Net. | 489/489 | 489/489 | 458/489 | NA |

Table 3: Number of stocks out of 489 total stocks where Model 1 has a lower out-of-sample error than Model 2: $\frac{1}{489}\sum_{j=1}^{489} \mathbf{1}_{\varepsilon^j_{\text{Model 1}} < \varepsilon^j_{\text{Model 2}}}$. "Neural Net." is the standard neural network architecture described in Section 4.1. "Spatial Neural Net." is the computationally efficient neural network architecture for spatial distributions developed in Section 4.3. Results are for the marginal distribution of the best ask price at a 1 second time horizon.

| Model 1/Model 2 | Naive empirical model | Logistic Reg. | Neural Net. | Spatial Neural Net. |
|---|---|---|---|---|
| Naive empirical model | NA | -5.94 | -9.63 | -10.31 |
| Logistic Reg. | 5.52 | NA | -3.51 | -4.14 |
| Neural Net. | 8.71 | 3.36 | NA | -0.62 |
| Spatial Neural Net. | 9.27 | 3.95 | 0.61 | NA |

Table 4: Average percent decrease in out-of-sample error for Model 1 versus Model 2: $\frac{1}{489}\sum_{j=1}^{489} \frac{\varepsilon^j_{\text{Model 2}} - \varepsilon^j_{\text{Model 1}}}{\varepsilon^j_{\text{Model 2}}} \times 100\%$. "Neural Net." is the standard neural network architecture described in Section 4.1. "Spatial Neural Net." is the computationally efficient neural network architecture for spatial distributions developed in Section 4.3. Results are for the marginal distribution of the best ask price at a 1 second time horizon.

| Model 1/Model 2 | Naive empirical model | Logistic Reg. | Neural Net. | Spatial Neural Net. |
|---|---|---|---|---|
| Naive empirical model | NA | 199/489 | 1/489 | 0/489 |
| Logistic Reg. | 290/489 | NA | 1/489 | 0/489 |
| Neural Net. | 488/489 | 488/489 | NA | 31/489 |
| Spatial Neural Net. | 489/489 | 489/489 | 458/489 | NA |

Table 5: Number of stocks out of 489 total stocks where Model 1 has a lower out-of-sample error than Model 2. "Neural Net." is the standard neural network architecture described in Section 4.1. "Spatial Neural Net." is the computationally efficient neural network architecture for spatial distributions developed in Section 4.3. Results are for the joint distribution of the best ask and best bid prices at a 1 second time horizon.

| Model 1/Model 2 | Naive empirical model | Logistic Reg. | Neural Net. | Spatial Neural Net. |
|---|---|---|---|---|
| Naive empirical model | NA | -0.58 | -14.63 | -15.36 |
| Logistic Reg. | 0.47 | NA | -14.01 | -13.74 |
| Neural Net. | 12.51 | 12.05 | NA | -0.64 |
| Spatial Neural Net. | 13.07 | 11.87 | 0.63 | NA |

Table 6: Average percent decrease in out-of-sample error for Model 1 versus Model 2. "Neural Net." is the standard neural network architecture described in Section 4.1. "Spatial Neural Net." is the computationally efficient neural network architecture for spatial distributions developed in Section 4.3. Results are for the joint distribution of the best ask and best bid prices at a 1 second time horizon.

## 6.2 Case [2]: Random Time Horizon at the Next Change of Bid or Ask Prices

Out-of-sample performance of the models is reported for the prediction case of the next change of the best bid or best ask prices. The models seek to predict the joint distribution of the best ask and best bid prices upon the *next price move*. The next price move is defined as the first time at which the best bid price or best ask price changes.

Tables 7 and 8 compare the out-of-sample performance of the different models for the marginal distribution of the best ask price for Case [2]. Tables 9 and 10 compare the out-of-sample performance of the different models for the *joint distribution* of the best ask and best bid prices for Case [2]. The neural networks consistently have lower error than the naive empirical model and the logistic regression. The spatial neural network consistently has lower error than the standard neural network.

| Model 1/Model 2 | Naive empirical model | Logistic Reg. | Neural Net. | Spatial Neural Net. |
|---|---|---|---|---|
| Naive empirical model | NA | 4/489 | 0/489 | 0/489 |
| Logistic Reg. | 485/489 | NA | 3/489 | 0/489 |
| Neural Net. | 489/489 | 486/489 | NA | 17/489 |
| Spatial Neural Net. | 489/489 | 489/489 | 472/489 | NA |

Table 7: Number of stocks out of 489 total stocks where Model 1 has a lower out-of-sample error than Model 2: $\frac{1}{489} \sum_{j=1}^{489} \mathbf{1}_{\varepsilon^j_{\text{Model 1}} < \varepsilon^j_{\text{Model 2}}}$. "Neural Net." is the standard neural network architecture described in Section 4.1. "Spatial Neural Net." is the computationally efficient neural network architecture for spatial distributions developed in Section 4.3. Results are for the marginal distribution of the best ask price at the time of the next price move.

| Model 1/Model 2 | Naive empirical model | Logistic Reg. | Neural Net. | Spatial Neural Net. |
|---|---|---|---|---|
| Naive empirical model | NA | -13.27 | -29.11 | -31.66 |
| Logistic Reg. | 11.56 | NA | -14.05 | -16.29 |
| Neural Net. | 22.03 | 11.80 | NA | -1.99 |
| Spatial Neural Net. | 23.54 | 13.51 | 1.92 | NA |

Table 8: Average percent decrease in out-of-sample error for Model 1 versus Model 2: $\frac{1}{489} \sum_{j=1}^{489} \frac{\varepsilon^j_{\text{Model 2}} - \varepsilon^j_{\text{Model 1}}}{\varepsilon^j_{\text{Model 2}}} \times 100\%$. "Neural Net." is the standard neural network architecture described in Section 4.1. "Spatial Neural Net." is the computationally efficient neural network architecture for spatial distributions developed in Section 4.3. Results are for the marginal distribution of the best ask price at the time of the next price move.

| Model 1/Model 2 | Naive empirical model | Logistic Reg. | Neural Net. | Spatial Neural Net. |
|---|---|---|---|---|
| Naive empirical model | NA | 41/489 | 0/489 | 0/489 |
| Logistic Reg. | 448/489 | NA | 2/489 | 0/489 |
| Neural Net. | 489/489 | 487/489 | NA | 16/489 |
| Spatial Neural Net. | 489/489 | 489/489 | 473/489 | NA |

Table 9: Number of stocks out of 489 total stocks where Model 1 has a lower out-of-sample error than Model 2. "Neural Net." is the standard neural network architecture described in Section 4.1. "Spatial Neural Net." is the computationally efficient neural network architecture for spatial distributions developed in Section 4.3. Results are for the joint distribution of the best ask and best bid prices at the time of the next price move.

| Model 1/Model 2 | Naive empirical model | Logistic Reg. | Neural Net. | Spatial Neural Net. |
|---|---|---|---|---|
| Naive empirical model | NA | -17.18 | -53.16 | -59.12 |
| Logistic Reg. | 13.00 | NA | -32.63 | -37.80 |
| Neural Net. | 32.47 | 21.38 | NA | -3.88 |
| Spatial Neural Net. | 34.83 | 24.13 | 3.50 | NA |

Table 10: Average percent decrease in out-of-sample error for Model 1 versus Model 2. "Neural Net." is the standard neural network architecture described in Section 4.1. "Spatial Neural Net." is the computationally efficient neural network architecture for spatial distributions developed in Section 4.3. Results are for the joint distribution of the best ask and best bid prices at the time of the next price move.

## 6.3 Model Performance in the Tail of the Distribution

This section examines model performance in the tail of the distribution. Specifically, we compare model performance for predicting the marginal distribution of the best ask price conditional on the best ask price increasing (the upper tail of the distribution). We first compare the cross-entropy error for the different models in the tail of the distribution. We also look at another metric for model performance, the *accuracy*. We discuss this metric in Section 6.3.1. We then present results for the out-of-sample model accuracies in the tail of the distribution in Section 6.3.2. The neural networks significantly outperform logistic regression and the naive empirical model for tail accuracy. The spatial neural network outperforms the standard neural network for tail accuracy.

Tables 11 and 12 compare the cross-entropy error of the different models in the *upper tail* of the distribution for Case [1]. Note that the outperformance of the spatial neural network increases substantially in the tail of the distribution compared to results for the full distribution in Section 6.1. As mentioned earlier, this matches the outperformance observed in Case [2].

| Model 1/Model 2 | Naive empirical model | Logistic Reg. | Neural Net. | Spatial Neural Net. |
|---|---|---|---|---|
| Naive empirical model | NA | 8/489 | 1/489 | 1/489 |
| Logistic Reg. | 481/489 | NA | 3/489 | 0/489 |
| Neural Net. | 488/489 | 486/489 | NA | 7/489 |
| Spatial Neural Net. | 488/489 | 489/489 | 482/489 | NA |

Table 11: Number of stocks out of $489$ total stocks where Model 1 has a lower out-of-sample error than Model 2: $\frac{1}{489} \sum_{j=1}^{489} \mathbf{1}_{\varepsilon^j_{\text{Model 1}} < \varepsilon^j_{\text{Model 2}}}$. "Neural Net." is the standard neural network architecture described in Section 4.1. "Spatial Neural Net." is the computationally efficient neural network architecture for spatial distributions developed in Section 4.3. Results are for the marginal distribution of the best ask price at a 1 second time horizon *conditional on the best ask price increasing*.

| Model 1/Model 2 | Naive empirical model | Logistic Reg. | Neural Net. | Spatial Neural Net. |
|---|---|---|---|---|
| Naive empirical model | NA | -20.92 | -28.90 | -34.03 |
| Logistic Reg. | 16.64 | NA | -6.51 | -10.68 |
| Neural Net. | 21.45 | 5.92 | NA | -3.86 |
| Spatial Neural Net. | 24.30 | 9.30 | 3.64 | NA |

Table 12: Average percent decrease in out-of-sample error for Model 1 versus Model 2: $\frac{1}{489} \sum_{j=1}^{489} \frac{\varepsilon^j_{\text{Model 2}} - \varepsilon^j_{\text{Model 1}}}{\varepsilon^j_{\text{Model 2}}} \times 100\%$. "Neural Net." is the standard neural network architecture described in Section 4.1. "Spatial Neural Net." is the computationally efficient neural network architecture for spatial distributions developed in Section 4.3. Results are for the marginal distribution of the best ask price at a 1 second time horizon *conditional on the best ask price increasing*.

### 6.3.1 Another metric: Accuracy

The cross-entropy error (i.e., the negative log-likelihood) is the best metric to evaluate model performance since it measures how well the model fits the empirical distribution of the data. However, it does lack some intuition in the sense that it is unclear how practically significant a reduction of $1\%$ in cross-entropy error is. A more interpretable metric is the *accuracy* of the model. Model accuracy is the percentage of the time where the model correctly predicts the outcome. The predicted outcome is taken as the most likely event according to the model-produced distribution. In this section and Section 6.3.2, we report the accuracy of neural networks in the limit order book setting.

In some settings, such as image classification, accuracy is an extremely good metric which in practice closely coincides with the cross-entropy error. However, in general, this may not be the case and we caution that accuracy may be an imperfect metric for many financial applications. Financial applications typically have a large amount of noise; modeling the distribution of the noise is just as important as modeling the most likely outcome. A simple example is the prediction of the binary event $Y \in \{0, 1\}$ where the true probability of event $Y = 1$ is $\frac{99}{100}$. The two models $\mathbb{P}[Y = 1] = \frac{51}{100}$ and $\mathbb{P}[Y = 1] = \frac{98}{100}$ both have the same accuracy (99%). However, the second model is clearly superior and has a much smaller cross-entropy error. Nonetheless, accuracy is an easily interpreted metric and thus can be worthwhile examining.

Figures 8 and 9 compare the out-of-sample accuracy of the naive empirical model, logistic regression, and standard neural network for the best ask price in Case [2]. The accuracies are for the full marginal distribution of the best ask price (there is no conditioning on the best ask price increasing or decreasing). Figures 8 and 9 are histograms for the increase in accuracy of the standard neural network over the naive model and logistic regression, respectively. Accuracies are measured in percent (i.e., if the neural network has an accuracy of 60% and the naive model has an accuracy of 51%, the increase in accuracy is 9%). The neural network offers signficant improvement over both logistic regression and the naive empirical model.

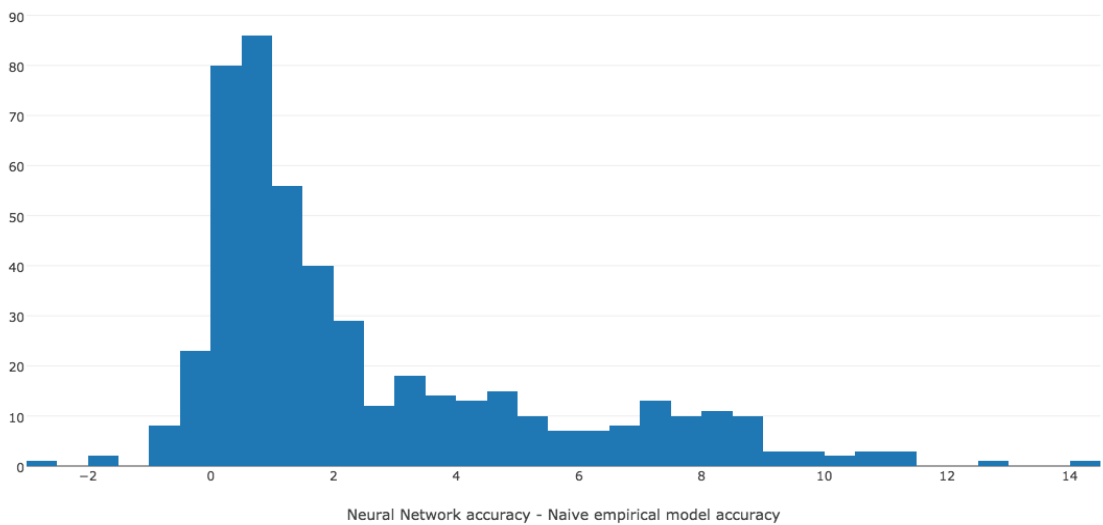

Figure 8: Increase in out-of-sample accuracy of neural network over naive empirical model. Accuracies are measured in percent. Results are for the marginal distribution of the best ask price at the time of the next price move.

### 6.3.2 Accuracy in the Tail of the Distribution

The difference between the standard neural network and the spatial neural network is in the tail of the distribution. We examine the *top-k accuracy* for the upper tail of the marginal distribution of the best ask price in Case [1]. We define a model's top-$k$ accuracy as the percent of time the actual outcome is in the model's top $k$ most likely outcomes. The top-1 accuracy is simply the model's accuracy. For each model, Table 13 reports the out-of-sample top-$k$ accuracy for the best ask price conditional on the best ask price increasing. The spatial neural network outperforms the standard neural network. The neural networks strongly outperform the logistic regression and naive models. Table 14 directly compares the top-$k$ accuracy of the spatial neural network and the standard neural network by reporting the fraction of stocks where the spatial neural network's top-$k$ accuracy is greater than the standard neural network's top-$k$ accuracy conditional on the best ask price increasing. Table 15 compares the top-$k$ accuracy of the spatial neural network and the logistic regression by reporting the fraction of stocks where the spatial neural network's top-$k$ accuracy is greater than the standard neural network's top-$k$ accuracy conditional on the best ask price increasing.

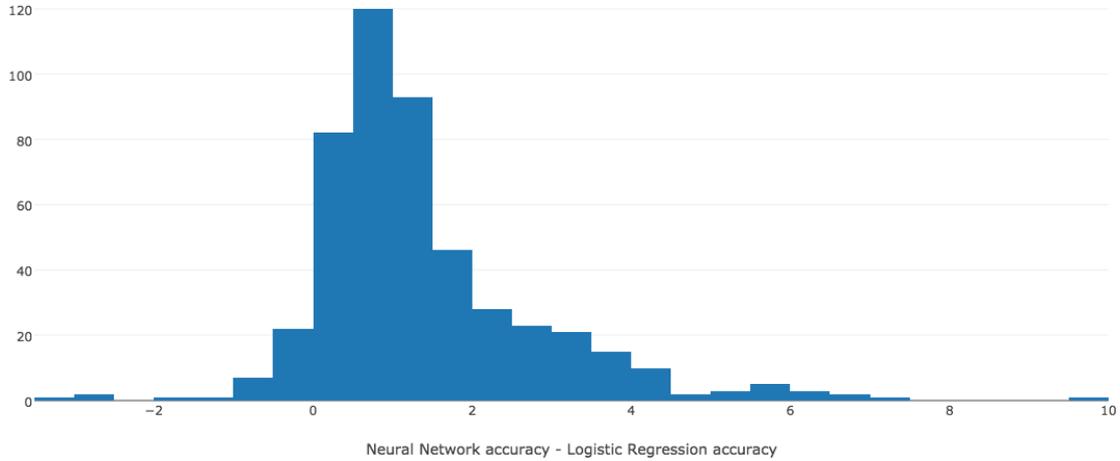

Figure 9: Increase in out-of-sample accuracy of neural network over logistic regression. Accuracies are measured in percent. Results are for the marginal distribution of the best ask price at the time of the next price move.

| k/Model | Naive empirical model | Logistic Regression | Neural Net. | Spatial Neural Net. |
|---|---|---|---|---|
| 1 | 62.04 | 66.42 | 69.90 | 70.98 |
| 2 | 78.96 | 82.97 | 84.92 | 86.15 |
| 3 | 86.26 | 89.44 | 90.68 | 91.77 |
| 4 | 90.31 | 92.68 | 93.56 | 94.53 |
| 5 | 93.13 | 94.55 | 95.23 | 96.09 |
| 6 | 94.44 | 95.67 | 96.21 | 97.00 |
| 7 | 95.36 | 96.42 | 96.90 | 97.62 |
| 8 | 96.06 | 96.95 | 97.41 | 98.07 |
| 9 | 96.61 | 97.34 | 97.80 | 98.40 |
| 10 | 97.06 | 97.64 | 98.10 | 98.65 |

Table 13: Average out-of-sample top-$k$ accuracy (in %). Top-$k$ accuracy is the percent of time the actual outcome is in the model's top $k$ most likely outcomes. Results are for the marginal distribution of the best ask price at a 1 second time horizon *conditional on the best ask price increasing*.

| k | Spatial Neural Net. vs. Neural Net. |
|---|---|
| 1 | 460/489 |
| 2 | 471/489 |
| 3 | 470/489 |
| 4 | 467/489 |
| 5 | 461/489 |
| 6 | 454/489 |
| 7 | 456/489 |
| 8 | 453/489 |
| 9 | 455/489 |
| 10 | 458/489 |

Table 14: Number of stocks out of $489$ total stocks where the spatial neural network's out-of-sample top-$k$ accuracy is greater than the standard neural network's out-of-sample top-$k$ accuracy. Top-$k$ accuracy is the percent of time the actual outcome is in the model's top $k$ most likely outcomes. Results are for the marginal distribution of the best ask price at a 1 second time horizon *conditional on the best ask price increasing*.

## 6.4 Error versus Computational Cost

Besides having lower error, the spatial neural network can learn more quickly than the standard neural network during training due to the reasons discussed in Sections 3 and 4. Figure 10 shows the out-of-sample error versus computational cost for the marginal distribution of the best ask price of Amazon in Case [1]. The computational time only includes the training time and does not include data pre-processing time. For Amazon, the standard neural network takes about 1700 seconds to reach its lowest error while the spatial neural network can achieve the same error in about 90 seconds.

| k | Spatial Neural Net. vs. Logistic Reg. |
|---|---|
| 1 | 487/489 |
| 2 | 487/489 |
| 3 | 486/489 |
| 4 | 485/489 |
| 5 | 485/489 |
| 6 | 486/489 |
| 7 | 488/489 |
| 8 | 484/489 |
| 9 | 488/489 |
| 10 | 486/489 |

Table 15: Number of stocks out of $489$ total stocks where the spatial neural network's out-of-sample top-$k$ accuracy is greater than the logistic regression's out-of-sample top-$k$ accuracy. Top-$k$ accuracy is the percent of time the actual outcome is in the model's top $k$ most likely outcomes. Results are for the marginal distribution of the best ask price at a 1 second time horizon *conditional on the best ask price increasing*.

Shorter training times are highly desirable. A financial institution would train models for a large number of stocks (for instance, the Wilshire 5000 contains 3,671 stocks). As new data arrives, models need to be re-trained with the new data. This would occur on a frequent basis (daily, weekly, or monthly). Neural network performance can be further improved by building *model ensembles* (often referred to as bagging). An ensemble is simply the average of many independently trained models. Each neural network (due to its random initial parameters and the random sequence of minibatches) will typically reach a different local minimum during training. By averaging the independently trained models, the variance can be significantly reduced and the out-of-sample error will decrease. If a financial institution builds an ensemble of neural networks for each stock, computational cost can further increase by an order of magnitude or more.

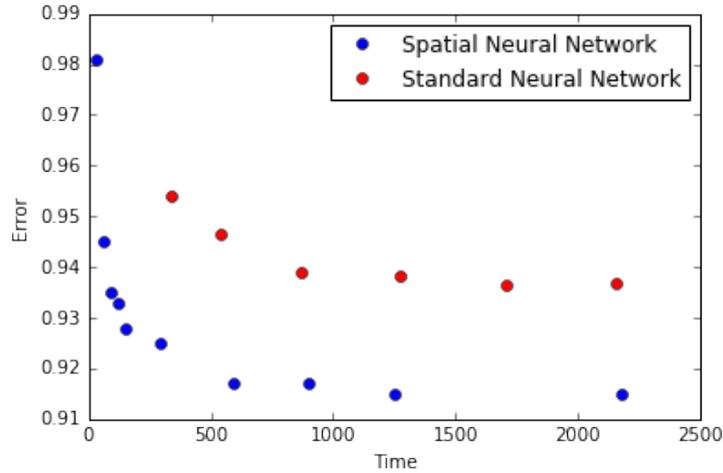

Figure 10: Out-of-sample error versus computational cost for Amazon.

# 7 Conclusion

This paper explores neural networks and deep learning for limit order book modeling. Neural networks are found to perform well for modeling the distribution of the best ask and best bid prices, with significantly better performance relative to logistic regression (with nonlinear features). The strong outperformance of the neural networks over logistic regression suggests that current industry risk modeling and management approaches can potentially be improved by adopting neural networks. Models are trained and tested using a dataset of nearly 500 U.S. stocks over a 20 month period.

Besides testing a standard neural network architecture, this paper develops a new neural network architecture (which we have referred to as a "spatial neural network") for modeling spatial distributions (i.e., distributions on $\mathbb{R}^d$). This new architecture has several advantages over the standard neural network architecture for modeling distributions on $\mathbb{R}^d$, including better generalization over space, lower computational expense, and the ability to take advantage of any local spatial structure. For the dataset considered in this paper, this spatial neural network has lower out-of-sample error, lower computational cost, and greater interpretability than the standard neural network. The spatial neural network's outperformance of the standard neural network can be largely attributed to its taking advantage of the limit order book's local spatial structure. The spatial neural network's architecture mimics this local behavior, yielding a low-dimensional model of movements deep into the limit order book. The spatial neural network performs especially well in the tail of the distribution, which is important for risk management applications.

This paper models the *joint distribution* of the best bid and best ask prices. This is essential for risk management applications (e.g., computing Value at Risk). This paper's approach could also be easily used to model the joint distribution of the best bid, best ask, and other limit order book features. Finally, although this paper focuses on limit order books, the spatial neural network provides benefits for any setting which requires modeling a distribution on $\mathbb{R}^d$.

# A  List of Stocks

LLTC, JD, NVDA, COG, BBBY, HAS, BRK.B, AES, ADT, HRS, GILD, ABBV, BA, ALXN, ALKS, HOG, BCR, AAL, CTSH, HON, AIZ, INTC, YHOO, HOLX, ILMN, INCY, BBT, BXP, CAM, BIIB, KLAC, ABT, LVNTB, HRL, BSX, CHKP, AGN, NXPI, BAX, CVC, BBY, CHTR, TSLA, HPQ, BHI, APD, IBKR, HCA, CBS, ISRG, ALLE, CTXS, BWA, ACE, ACN, ADBE, ADP, CCL, GWW, CPB, BMY, HIG, CA, AIV, HAL, BRCM, ADS, CBG, BDX, SCTY, LMCA, AA, HSIC, AMGN, ADI, AMZN, HAR, ADM, AVY, NFLX, MXIM, CAT, MSFT, ENDP, MAR, COF, AAP, NTES, BLL, FOXA, AEE, DISH, CTRP, KMX, AMAT, HCP, HBI, CHRW, BK, ATVI, BAC, ADSK, AAPL, CAH, HRB, GOOGL, BMRN, FB, GT, MKC, QCOM, MET, NDAQ, SBUX, STT, NOV, MYL, SYF, TGT, JWN, SNI, PNR, MHFI, PH, TEL, MCD, NLSN, MCO, MS, MUR, ORCL, SYK, SLB, PCAR, PCL, SO, OI, PPG, NFX, NI, SIG, SE, PRGO, SPLS, TAP, PKI, RL, PBI, PDCO, PXD, MCHP, PCG, ROST, MCK, SEE, PSX, MON, SWN, R, STI, MOS, RCL, SLG, CRM, PPL, MRK, MAT, MSI, PNC, PBCT, SCG, NEM, PNW, PM, PEP, SNA, NKE, SYY, MA, TE, NTAP, NAVI, SHW, PFE, POM, SWK, MJN, SPG, COL, SRE, ROP, HOT, MNST, SJM, STJ, PAYX, NEE, NWSA, OKE, MHK, MDT, NWL, CSC, DISCA, LEG, DVN, HES, PGR, OMC, MMC, COH, DUK, JPM, DG, CNP, DFS, GMCR, MAC, CCI, CMG, CVS, MTB, DRI, CCE, DE, ICE, MAS, DGX, PEG, CSX, JBHT, DNB, COP, COST, DD, MPC, DISCK, DHR, DLTR, CLX, CELG, CL, CMS, CB, LB, PG, KR, HD, LH, HBAN, PFG, LLL, PVH, CMA, KEY, HP, KIM, CVX, PWR, LUK, OXY, DHI, JNPR, D, ITW, DOV, DO, CME, CMI, LEN, PHM, DPS, KSU, CMCSA, DOW, O, KMB, PRU, CTAS, PLD, MRO, LOW, DTE, CINF, L, M, K, C, CHK, LNC, CF, LM, CI, MLM, HST, CERN, LVLT, DAL, LYB, LLY, JNJ, KSS, CAG, KMI, IR, PSA, CTL, DLPH, LMT, DVA, IBM, ATML, TMO, NRG, TXN, UAL, YUM, TWX, HSY, HCN, WEC, VAR, UHS, VLO, NSC, WYN, FLS, WDC, UNH, FLIR, WU, FE, THC, ZTS, ZION, FIS,

TSO, TRV, RIG, ECL, A, WM, TIF, WAT, VFC, XRX, WFC, FOSL, WMB, FCX, NUE, TSN, USB, EA, WBA, XYL, TSS, NTRS, WYNN, WY, UNM, UA, XL, VZ, TXT, BEN, PCLN, TWC, ED, URI, UTX, FLR, TDC, UPS, NOC, EL, NVAX, UNP, FISV, VMC, FDX, WMT, VNO, DIS, TMK, V, XEL, REGN, FITB, TJX, WFM, TYC, VTR, XEC, FFIV, SCHW, EIX, WHR, URBN, FAST, GRPN, FSLR, VIAB, ETN, RTN, JBLU, VRSN, EXPD, EQT, ETR, KO, EMC, SYMC, AKAM, GD, EXPE, SWKS, DOX, JAZZ, STX, RRC, FOX, EQIX, XLNX, IPG, GE, SBAC, GME, LRCX, EMR, CSCO, RF, TSCO, TRIP, MU, QVCA, XOM, TROW, GRMN, IP, JCI, ESRX, GPS, INTU, ROK, MDLZ, FTI, EMN, ESS, LULU, GS, SNDK, JEC, SHPG, PX, RSG, BIDU, VOD, RAI, EXC, VRSK, EW, GIS, FTR, SRCL, IRM, GM, ORLY, EQR, F, ETFC, RHT, IFF, HUM, GGP, RHI, LKQ, VIA, TMUS, SIRI, ULTA, AVGO, IVZ, VRTX, EOG, EBAY, NCLH, XRAY, GPC

## B  Bounded Hidden Units

Without loss of generality let, $\mathcal{R}_+ = \{1, 2, \ldots\}$. Let $q_k = \mathbb{P}[Y = k | Y \geq k]$ for $k = 1, 2, \ldots$. If the hidden units are bounded, the output of the neural network $f_\theta(x, y)$ is bounded. Since $q_k$ is the softmax of a bounded function, $0 < a \leq q_k \leq b < 1$. Let $p_k = \mathbb{P}[Y = k] = q_k \prod_{i=1}^{k-1}(1 - q_i)$. Also, define:

$$F_N = \sum_{n=1}^{N} p_n. \tag{14}$$

We want to show that $F_N \to 1$ as $N \to \infty$. In other words, the distribution of $Y$ does not have positive mass at $+\infty$.

Let $\tilde{q}_k = q_k$ for $k \leq N$ and $\tilde{q}_{N+1} = 1$. Let $\tilde{p}_k = \tilde{q}_k \prod_{i=1}^{k-1}(1 - \tilde{q}_i)$. Consider a random variable $Z \in \mathcal{R}_+$ where $\mathbb{P}[Z = k | Z \geq k] = \tilde{q}_k$ and $\mathbb{P}[Z = k] = \tilde{p}_k$. $Z$ is a well-defined random variable (e.g., the sum of its probabilities is 1).

Note that $F_N$ can be rewritten as:

$$F_N = \sum_{n=1}^{N} p_n = \sum_{n=1}^{N} \tilde{p}_n = 1 - P[Z > N] = 1 - \prod_{n=1}^{N}(1 - \tilde{q}_n) = 1 - \prod_{n=1}^{N}(1 - q_n) \tag{15}$$

Now, we have that for all $N$:

$$1 - \prod_{n=1}^{N}(1-a) \leq F_N \leq 1 - \prod_{n=1}^{N}(1-b) \qquad (16)$$

The LHS and RHS converge to 1, which shows $F_N \to 1$.

Examples of bounded hidden units include tanh, sigmoid, and clipped ReLU. The above result also holds if at least one of the hidden layers has bounded units, but the units in the other hidden layers are unbounded (e.g., ReLU).

In general, if $q_k$ has no positive lower bound, $F_N$ may not converge to 1. This is due to $Y$ potentially escaping to $+\infty$ with positive probability. An extreme example is if $q_k = 0$. ReLU units are not bounded. Consequently, if the last hidden layer has ReLU units, $q_k$ is not bounded from below. We analyze the case of ReLU units in Section C of the Appendix.

## C  ReLU Hidden Units

Rectified linear units are of the form $\max(z, 0)$ for an input $z$. We re-state the equation for $F_N$ from Section B:

$$F_N = 1 - \prod_{n=1}^{N}(1 - q_n). \qquad (17)$$

Recall that

$$1 - q_n = \frac{1}{1 + e^{f_\theta(x,n)}}, \qquad (18)$$

where $f_\theta$ is a neural network with ReLU hidden units. If all hidden units of the neural network $f_\theta(x, y)$ are ReLU units, $f_\theta(x, y)$ has three possible forms for large $y$: (1) 0, (2) $C_2 + K_2 y$, or (3) $C_3 - K_3 y$. More precisely, there exists an $N_0$ such that $f_\theta(x, y)$ equals either (1), (2), or (3) for all $y \geq N_0$. The specific form the neural network takes for large $y$ depends upon the parameters $\theta$. $F_N \to 1$ as $N \to \infty$ for cases (1) and (2). However, $F_N \to \bar{F}$ where $0 < \bar{F} < 1$ in case (3).

To prove that $F_N \to \bar{F}$ where $0 < \bar{F} < 1$ in case (3), we will show an upper bound $F_N \leq G_N$ where $G_N \to G < 1$. Note that $F_N$ has a limit $\bar{F}$ since it is bounded ($0 \leq F_N \leq 1$) and monotone increasing;

however, this limit will not be 1.

$$F_N = 1 - \prod_{n=1}^{N} \frac{1}{1 + e^{f_\theta(x,n)}} = 1 - \exp\big(-\sum_{n=1}^{N} \log(1 + e^{f_\theta(x,n)})\big). \quad (19)$$

Using the inequality $\log(1+z) \leq z$ for $z > -1$:

$$\sum_{n=1}^{N} \log(1 + e^{f_\theta(x,n)}) \leq \sum_{n=1}^{N} e^{f_\theta(x,n)} \leq D_3 + \sum_{n=N_0}^{N} e^{f_\theta(x,n)}$$

$$= D_3 + \sum_{n=N_0}^{N} e^{C_3 - K_3 n} \leq D_3 + e^{C_3 - K_3 N_0} + \int_{n=N_0}^{N} e^{C_3 - K_3 y} dy. \quad (20)$$

The RHS of (20) converges to a finite positive number as $N \to \infty$. Combining (20) with (19) implies that:

$$F_N = 1 - \exp\big(-\sum_{n=1}^{N} \log(1 + e^{f_\theta(x,n)})\big) \leq G < 1. \quad (21)$$

Therefore, in case (3), $F_N$ does not converge to 1 as $N \to \infty$.

We now show that $F_N \to 1$ in cases (1) and (2). In case (1),

$$\sum_{n=1}^{N} \log(1 + e^{f_\theta(x,n)}) = D_1 + \sum_{n=N_0}^{N} e^0 \underset{N \to \infty}{\to} \infty \quad (22)$$

Combining (22) with (19) implies that $F_N \to 1$ as $N \to \infty$. For case (2), consider the lower bound:

$$\sum_{n=1}^{N} \log(1 + e^{f_\theta(x,n)}) \geq \sum_{n=1}^{N} f_\theta(x,n) = D_2 + \sum_{n=N_0}^{N} (C_2 + K_2 n) \underset{N \to \infty}{\to} \infty. \quad (23)$$

Combining (23) with (19) (and recalling that $F_N \leq 1$) implies that $F_N \to 1$ as $N \to \infty$. Note that the constants $D_1, D_2, D_3, C_2$, and $C_3$ implicitly depend upon $x$ and $\theta$.

## D  Comparison against a theoretical model

Theoretical models have made key contributions to the economic understanding of limit order book dynamics. For instance, the formula in Cont & Larrard (2012) reveals that limit order book dynamics intrinsically depend upon the order book imbalance, which is a nonlinear feature widely used in trading algorithms and as an input to statistical models. However, data-driven models may provide better quantitative predictions than theoretical models in practice.

The theoretical models cannot be directly compared against the data-driven models in this paper due to the theoretical models' lack of tractability for computing a distribution on $\mathbb{N} \times \mathbb{N}$, which is the quantity this paper is interested in. For the interested reader, we provide here a comparison in the much simpler case of predicting the direction of the next move of the best ask price (i.e., the probability that it moves up or down). We compare the model from the seminal paper of Cont & Larrard (2012) against logistic regression and a neural network. The model from Cont & Larrard (2012) for the direction of the next price move is

$$p^{\text{up}} = \frac{1}{2} - \frac{\arctan(\sqrt{\frac{1+\rho}{1-\rho}} \frac{\text{best ask size} - \text{best bid size}}{\text{best ask size} + \text{best bid size}})}{2 \arctan(\sqrt{\frac{1+\rho}{1-\rho}})}, \tag{24}$$

where $\rho$ is the correlation between the increments of the best bid and best ask sizes. Tables 16 and 17 compare the data-driven models with the theoretical model. The comparison is performed using a subset of 109 stocks from the dataset. The "error" is the cross-entropy error (i.e., the negative log-likelihood).

| Model 1/Model 2 | Logistic Reg. | Neural Net. | Theoretical |
|---|---|---|---|
| Logistic Reg. | NA | 1/109 | 109/109 |
| Neural Net. | 108/109 | NA | 109/109 |
| Theoretical | 0/109 | 0/109 | NA |

Table 16: Number of stocks out of 109 total stocks where Model 1 has a lower out-of-sample error than Model 2: $\frac{1}{109} \sum_{j=1}^{109} \mathbf{1}_{\varepsilon^j_{\text{Model 1}} < \varepsilon^j_{\text{Model 2}}}$. Results are for the marginal distribution of the *direction* (i.e., up or down) of the next change of the best ask price. A subset of 109 stocks is used.

| Model 1/Model 2 | Logistic Reg. | Neural Net. | Theoretical |
|---|---|---|---|
| Logistic Reg. | NA | -64.58 | 25.83 |
| Neural Net. | 28.20 | NA | 43.92 |
| Theoretical | -59.69 | -201.76 | NA |

Table 17: Average percent decrease in out-of-sample error for Model 1 versus Model 2: $\frac{1}{109} \sum_{j=1}^{109} \frac{\varepsilon^j_{\text{Model 2}} - \varepsilon^j_{\text{Model 1}}}{\varepsilon^j_{\text{Model 2}}} \times 100\%$. Results are for the marginal distribution of the *direction* (i.e., up or down) of the next change of the best ask price. A subset of 109 stocks is used.